\documentclass[preprint]{aastex}
\usepackage{epsfig}
\usepackage{lscape}
\usepackage{graphicx}
\usepackage{latexsym}
\usepackage{color}

\begin{document}

\title{Constructing Synchronously Rotating Double White Dwarf Binaries}

\author{Wesley Even and Joel E. Tohline}
\affil{Department of Physics and Astronomy, Louisiana State
University,
    Baton Rouge, LA 70803}

\begin{abstract}
We have developed a self-consistent-field technique similar to the one described by \cite{hachisuEtAl86b} that can be used to
construct detailed force-balanced models of synchronously
rotating, double white dwarf (DWD) binaries that have a wide range of total masses,
mass ratios, and separations.  In addition to providing a
computational tool that can be used to provide quiet initial
starts for dynamical studies of the onset of mass transfer in
DWD systems, we show that this SCF technique can be used to
construct model sequences that mimic the last portion of the
detached inspiral phase of DWD binary evolutions, and
semi-detached model sequences that mimic a phase of
conservative mass transfer.

\end{abstract}


\section{Introduction}

As \cite{Webbink84} and \cite{IT84,IT86} have pointed out,
double white dwarf (DWD) binaries are expected to be the end
product of the thermonuclear evolution of a sizeable fraction
of all binary systems.  The subset of DWD binaries that are
born with orbital periods $P_\mathrm{orb} \lesssim 5$ hours are
of particular astrophysical interest because they will be
driven into contact within a Hubble time via the loss of
angular momentum through gravitational radiation \citep{Pac67}.
Even this short-period component of the DWD population is
expected to be quite large in our Galaxy. Sufficiently large
that, according to \cite{HBW90} and \cite{CL03}, DWDs are
likely to be a dominant source of background noise for the
proposed {\it Laser Interferometer Space Antenna} ({\it LISA})
\citep{FB84,EIS87,Bender98} in its lower gravitational-wave
frequency band, $f_\mathrm{GW} = 2/P_\mathrm{orb} \lesssim
3\times 10^{-3}$ Hz.  Ironically, it is difficult to detect
DWDs via traditional observational techniques because the
intrinsic photon luminosity of white dwarfs is very low.

Broadly speaking, our Galaxy's DWD binary population should be
dominated by systems that are in two distinctly different
evolutionary phases:  the inspiral phase alluded to above,
during which both stars are detached from their respective
Roche lobes; and a semi-detached, stable mass transfer phase,
during which the less massive star fills its Roche lobe and is
slowly transferring mass to its more massive companion. To date
$\sim 100$ detached systems have been identified
\citep{Nap01,Nap04}. Orbital periods and component masses have
been determined for approximately one quarter of this sample
\citep{Marsh00,MMM0a,Nap02,Karl03,NNKMV05}. AM CVn is the
prototype of semi-detached DWD systems that are undergoing a
phase of stable mass transfer \citep{Warner95}.  In the
immediate solar neighborhood $\sim 20$ such systems are known
\citep{WW03, Nelemans05, Anderson2005, Roelofs05, Ramsay07}.

Theoretical arguments suggest that the ultimate fate of a DWD
binary will depend on the system's total mass,
$M_\mathrm{tot}$, and mass ratio $q_0$ at the onset of mass
transfer \citep{MNS04, GPF07, Frank08}.  For example, a DWD
will likely only be able to enter an extended phase of stable
mass transfer as characterized by AM CVn systems if $q_0$ is
less than some critical value $q_\mathrm{stable}$
--- where $q_\mathrm{stable}$ is almost certainly $\le 2/3$ and
may be closer to $1/4$.  If $q_0 > q_\mathrm{stable}$, the mass
transfer rate is expected to diverge in a finite time,
ultimately implying tidal disruption of the donor and/or a
catastrophic merger of the two components.  Consistent with the
theoretical ideas presented by \cite{Webbink84} and
\cite{ITY96}, recent observations strongly suggest that the end
product of some DWD mergers are R Coronae Borealis (RCB) stars
and hydrogen-deficient carbon (HdC) stars \citep{CGHFA07}.
Also, DWD systems with $M_\mathrm{tot}$ in excess the
Chandrasekhar mass have long been considered likely progenitors
of Type Ia supernovae \citep{Webbink84, ITY96, livio00,
Yoon07}.

Over the past couple of decades, various groups have employed
smoothed particle hydrodynamics (SPH) techniques to illustrate
the dynamical behavior of DWD systems that violently merge
after encountering an unstable mass-transfer event
\citep{benz90, RS95, SCM97, FWHD99, GGBI04, Yoon07}. Typically,
initial states for these simulations have been constructed in
such a way that the merger process is completed in $\lesssim 5$
orbits after first contact. \cite{DMTF06} and \cite{MFTD07}
recently employed a grid-based finite-volume (FV) computational
fluid dynamic technique to also study the onset and nonlinear
development of mass transfer in strongly interacting binary
systems.  Their primary objective was to ascertain the value of
$q_\mathrm{stable}$ in systems, such as DWDs, that undergo a
phase of direct impact accretion following the onset of mass
transfer.  They were able to follow the evolution of a couple
of different systems through $\gtrsim 30$ orbits.  Instead of
merging, these systems appeared to be entering a long-term
phase of stable mass transfer.  Presumably these are the types
of binary configurations that serve as the progenitors of AM
CVn systems.

\cite{MFTD07} found that the outcome of their simulations ---
for example, whether a system survives the onset of mass
transfer or merges --- can be sensitive to initial conditions.
In particular if, rather than making only marginal contact with
its Roche lobe, the donor star is in relatively deep contact at
the onset of a simulation, the mass-transfer rate will
initially be artificially high and it may be difficult for the
system to avoid merger. As a result, a DWD system that should
be categorized by numerical simulation as an AM CVn progenitor
may be incorrectly categorized as a likely progenitor of an RCB
star or a Type Ia supernova.  It is therefore important to
start such simulations from initial states that are very quiet
--- that is, from initial configurations that are in detailed
force balance throughout --- and to perform each simulation
with a grid (or particle) resolution that is sufficient to
resolve marginal, or at least very shallow, contact between the
donor and its Roche lobe.  In two very recent reports,
\cite{FD08} and \cite{DRB08} have confirmed this finding.  Both
groups have shown that SPH techniques also can be used to
follow stable mass-transfer events through $\gtrsim 30$ orbits
if sufficiently quiet initial states are used and if the
simulations are carried out using a sufficiently large number
of SPH particles.

Quiet initial states were constructed for the hydrodynamic
simulations reported by \cite{DMTF06} and \cite{MFTD07} using a
self-consistent-field (SCF) technique very similar to the one
described by \cite{hachisu86b}.  However, these initial models were constructed
using a polytropic equation of state. Here we use a modified version
of the SCF technique developed by \cite{hachisuEtAl86b} for the zero temperature white dwarf (ZTWD) 
equation of state derived by \cite{Chandrasekhar35}.  We show how this SCF
technique can be used to construct synchronously rotating,
unequal-mass DWD binaries in which the less-massive (donor)
star is in marginal contact with its Roche lobe and thereby
provide excellent initial models for dynamical studies of
interacting DWD systems that have a realistic equation of
state.

To demonstrate the broad utility of this SCF technique,
we construct a sequence of detached, synchronously rotating
binaries of varying separation, but fixed mass ratio and
constant total mass, to mimic the last portion of the inspiral
phase of evolution of DWD binary systems. This enables us to
determine, for example, the degree to which the functional
dependance of the orbital frequency on orbital separation
$\Omega(a)$ deviates from a pure Keplerian behavior. This type
of sequence can be used to accurately identify the critical
separation $a_\mathrm{crit}$ and corresponding orbital
frequency at which the less massive component of a DWD binary
first makes contact with its Roche lobe. Information of this
nature will be helpful in interpreting future observations that
are expected to be made with {\it LISA}, as described for
example by \cite{KT07}. Finally, we show how this SCF
technique can be used to construct a sequence of semi-detached
binaries of fixed total mass, but varying separation and
decreasing mass ratio, to mimic the initial stage of evolution
of a DWD binary that has entered an AM CVn ({\it i.e.}, stable
mass-transfer) phase of evolution.

\section{SCF Formulation}
\label{sec:formulation}

The SCF technique was first introduced to the astrophysics
community by \cite{Ostriker64} to create models of rapidly
rotating, single stars with a polytropic equation of state.
Hachisu developed a variation of the technique, improving
convergence rates and extending its capabilities to include the
use of a ZTWD equation of state. With his improved technique,
Hachisu was able to construct two-dimensional (2D)
configurations of differentially rotating, single white dwarfs
\citep{hachisu86a} and three-dimensional (3D) configurations of
uniformly rotating multiple white dwarf systems in which the
stars have equal mass \citep{hachisu86b}.  \cite{New97}
employed Hachisu's 3D technique to construct inspiral sequences
of equal-mass DWD binaries, including over-contact models
having separations even smaller than $a_\mathrm{crit}$.
\cite{hachisu86b} also applied his technique to the
construction of unequal-mass binary systems using a polytropic
equation of state and after additional algorithmic innovations were introduced, Hachisu, Eriguchi, and Nomoto (1986a,b) constructed a small sample of  unequal-mass DWD binaries and heavy-disk white dwarf systems to examine the likely outcome of DWD mergers.  In what follows we show how Hachisu's SCF technique for constructing unequal-mass DWD binaries can be further improved and used to construct inspiral binary sequences.

\subsection{Equation of State}

In the zero-temperature white dwarf (ZTWD) equation of state
\citep{Chandrasekhar35, Chandrasekhar67, hachisu86a} the
electron degeneracy pressure $P$ varies with the mass density
$\rho$ according to the relation,
\begin{eqnarray}
P &=& A \biggl[ x(2x^2 - 3)(x^2 + 1)^{1/2} + 3 \sinh^{-1}x \biggr]
\, , \label{eq:EOSchandra}
\end{eqnarray}
where the dimensionless parameter,
\begin{eqnarray}
x &\equiv& \biggl(  \frac{\rho}{B} \biggr)^{1/3} \, , \label{eq:DefineX}
\end{eqnarray}
and the constants $A$ and $B$ are (see Appendix
\ref{Ap:MRrelationship} and Table \ref{tab:physicalConstants}
for elaboration),
\begin{eqnarray}
A &\equiv& \frac{\pi m_e^4 c^5}{3h^3} = 6.00228\times 10^{22}
~\mathrm{dynes}~\mathrm{cm}^{-2} \, ,
\\
\frac{B}{\mu_e} &\equiv& \frac{8\pi m_p }{3}\biggl( \frac{m_e c}{h} \biggr)^3
= 9.81011\times 10^5 ~\mathrm{g}~\mathrm{cm}^{-3} \, .
\end{eqnarray}
According to \cite{Chandrasekhar67} (see again our Appendix
\ref{Ap:MRrelationship}), a natural length scale associated
with models of ZTWDs is,
\begin{eqnarray}
\mu_e \ell_1 &=& \biggl( \frac{2A}{\pi G} \biggr)^{1/2} \frac{\mu_e}{B}
= 7.71395\times 10^8 ~\mathrm{cm} = 0.0111 R_\odot \, ,
\end{eqnarray}
and the associated limiting white dwarf mass is,
\begin{eqnarray}
 \mu_e^2 M_\mathrm{ch} &=& 4\pi (2.01824) \biggl(
\frac{2A}{\pi G} \biggr)^{3/2} \biggl( \frac{\mu_e}{B} \biggr)^2
= 1.14205 \times 10^{34}~\mathrm{g} = 5.742 M_\odot \, .
\end{eqnarray}
Throughout this work, we will assume that the average ratio of
nucleons to electrons throughout each white dwarf is $\mu_e =
2$. Hence, $B=1.96202 \times 10^6
~\mathrm{g}~\mathrm{cm}^{-3}$, $\ell_1 = 5.55 \times 10^{-3}
R_\odot$, and $M_\mathrm{ch} = 1.435 M_\odot$.

In terms of the enthalpy of the gas,\footnote{As defined here,
$H$ is actually enthalpy per unit mass.}
\begin{eqnarray}
H &\equiv& \int \frac{dP}{\rho} \, ,
\end{eqnarray}
the ZTWD equation of state shown in Eq.~(\ref{eq:EOSchandra})
can also be written in the form,
\begin{eqnarray}
H &=& \frac{8A}{B} \biggl[ x^2 + 1 \biggr]^{1/2} \, .
\label{eq:enthalpy01}
\end{eqnarray}
Inverting this gives the dependence of $\rho$ on $H$, namely,
\begin{eqnarray}
\frac{\rho}{B} &=& x^3 = \biggl[ \biggl( \frac{BH}{8A} \biggr)^2 - 1
\biggr]^{3/2} \, .
\end{eqnarray}

As a foundation for both constructing and understanding the
structures of the synchronously rotating and tidally distorted
stars in ZTWD binary systems, we have regenerated
Chandraskehar's spherical white dwarf sequence using a
variation of the SCF technique outlined by \cite{hachisu86a}.
As is discussed in \S \ref{sec:SingleWDs},  Table
\ref{tab:scf_single} details key properties of the ZTWD
structures that lie along this spherical model sequence. The
white dwarf mass-radius relationship that is derived from
models along this sequence is illustrated by the diamonds in
Figure \ref{fig:MR}.  For comparison, results from the
published spherical sequence of \cite{hachisu86a} are
represented in this figure by asterisks and the solid curve
shows the approximate, analytic mass-radius relationship,
Eq.~(\ref{eq:MRnauenbergAppendix}), derived for ZTWD stars by
\cite{Nauenberg72}.  (As explained in Appendix
\ref{Ap:MRrelationship}, it is more appropriate for us to
compare our results to this ``Nauenberg'' mass-radius relation
than to the more widely used ``Eggleton'' mass-radius relation,
shown in Eq.~\ref{eq:MReggleton}.)

\subsection{Binary System Geometry and Governing Equations}

Our objective is to determine
the 3D structure of a pair of ZTWD stars that are in a tight,
circular orbit under the condition that both stars are synchronously
rotating with the binary orbital frequency $\Omega$.
We begin by specifying the masses $M_1$ and $M_2$ of the primary and
secondary stars, respectively, such that $M_2 \leq M_1$.
Alternatively, we can specify the
total system mass $M_\mathrm{tot} \equiv M_1 + M_2$ and
the system mass ratio $q \equiv M_2/M_1 \leq 1$, in which case,
\begin{eqnarray}
M_1 &=& \biggl( \frac{1}{1+q}  \biggr) M_\mathrm{tot} \, , \nonumber \\
M_2 &=& \biggl( \frac{q}{1+q}  \biggr) M_\mathrm{tot} \, . \nonumber
\end{eqnarray}
Figure \ref{fig:diagram} shows a slice through the equatorial
plane of such a system under the assumption that both stars are
spherically symmetric. For our final equilibrium models in
which the effects of tidal and rotational distortions are taken
into account in a fully self-consistent fashion, this figure
provides only a schematic illustration of the binary system's
equatorial-plane structure. However, it provides an accurate
depiction of the equatorial-plane structure of the initial
stellar models that are fed into our iterative SCF scheme (see
\S \ref{sec:strategy}).

In Figure \ref{fig:diagram}, the more massive, primary star is
shown on the left and the less massive, secondary star is on
the right; the centers of the stars are located a distance
$\varpi_1$ and $\varpi_2$, respectively, from the center of
mass of the binary system; and the binary separation
$a=\varpi_1 + \varpi_2$. Because we are using a ZTWD equation
of state, the central density $\rho_\mathrm{max}^i$ and radius
$R_i$ of each star ($i=1,2$) cannot be specified independently
of each star's chosen mass. As an initial guess for our SCF
technique, the values of $\rho_\mathrm{max}^{i=1}$,
$\rho_\mathrm{max}^{i=2}$, $R_1$, and $R_2$ are drawn from
Table \ref{tab:scf_single}, that is, they are given by values
that correspond to spherical ZTWDs having masses $M_1$ and
$M_2$.

For various values of the three principal system parameters $M_1$,
$M_2$ and $a$, our specific aim is to determine in a self-consistent
fashion on a cylindrical coordinate mesh $(\varpi,\theta,Z)$, the
values and spatial distribution of the scalar fields
$\rho^i(\varpi,\theta,Z)$, $H^i(\varpi,\theta,Z)$, and
$P^i(\varpi,\theta,Z)$, for both stars ($i=1,2$) in synchronously
rotating, ZTWD binaries.  Following Hachisu (1986a,b) and Hachisu, Eriguchi, and Nomoto (1986a,b), in order to
construct these desired binary configurations we need to solve the
following five equations simultaneously:
\begin{eqnarray}
\nabla^2\Phi(\varpi,\theta,z) &=& 4\pi G \sum_i \rho^i(\varpi,\theta,z) \, ,
\label{eq:Poisson} \\
C^i &=& H^i(\varpi,\theta,z) + \Phi(\varpi,\theta,z) - \frac{1}{2}\Omega^2\varpi^2 \, ,
\label{eq:Bernoulli} \\
\rho^i(\varpi,\theta,z) &=& B[x^i(\varpi,\theta,z)]^3 =
B\biggl\{ \biggl[ \frac{B}{8A}~H^i(\varpi,\theta,z) \biggr]^2 -1
\biggr\}^{3/2} \, ,\label{eq:EOS1}
\end{eqnarray}
where $\Phi(\varpi,\theta,Z)$ is the Newtonian gravitational potential
of the combined stellar system, and $C^{i=1}$ and $C^{i=2}$ are
constants that specify the conditions of the Bernoulli flow inside
each star.

\subsection{Solution Strategy}\label{sec:strategy}
Our solution strategy follows closely the methods described in \cite{hachisu86b} and \cite{hachisuEtAl86b} so only the differences between our methods will be given in detail here.  The variables with a carat (\verb ^ ) above them are the dimensionless version of the variables as defined in Eqs.(22)-(27) of \cite{hachisu86b}.
 
To begin an SCF iteration two spherical stars are initially
placed on the computational grid in such a way that the center of
mass of the system falls at the origin of the coordinate system and
the outer edge of the secondary star (point {\bf $O_\alpha$} in
Figure \ref{fig:diagram}) is at $\hat{\varpi} = 1$.  The centers of
the stars are therefore located, respectively, at
\begin{eqnarray}
\hat{\varpi}_1 \equiv \frac{\varpi_1}{\varpi_\alpha} &=&
\frac{q}{1+\ell(1+q)} \, , \\
\hat{\varpi}_2 \equiv \frac{\varpi_2}{\varpi_\alpha} &=&
\frac{1}{1+\ell(1+q)} \, ,
\end{eqnarray}
where the dimensionless ratio $\ell \equiv R_2/a$ is known once
$M_2$ and $a$ have been chosen.  These two expressions make
sense because $\varpi_\alpha = (\varpi_2 + a\ell) = [\varpi_2 +
(\varpi_2 + \varpi_1)\ell]$ and, for a point-mass binary whose
center of mass is at the origin of the grid, $\varpi_1 = q
\varpi_2$.
With $\hat{\rho}^i$ defined everywhere on the grid,
$\hat{\Phi}(\hat\varpi,\theta,\hat{z})$ is calculated via
Eq.~(\ref{eq:Poisson}). In this work the boundary values for
$\hat{\Phi}$ are calculated using the compact cylindrical Green's
function expansion described in \cite{cohl99}, and the values of the
potential throughout the interior volume of the computational grid
are calculated using the Krylov subspace methods provided by the
PETSc software library \citep{PETSc}.

During each iteration, the interior structure of the secondary star is calculated using the same strategy as outlined in \S 2b of \cite{hachisuEtAl86b}.
To update the structure of the primary star, however, the location of $\hat{H}^{i=1}_{max}$ is used instead of specifying the inner edge of the star, as was done in \cite{hachisuEtAl86b}.  This choice eliminates the extra iterative steps that were needed when the third boundary condition was specified at the inner edge of the primary star.
Using Eq.~(\ref{eq:enthalpy01}), we determine the value of the
normalized enthalpy at the center of the primary star from the
values of $x^i_\mathrm{max}$ selected for both stars and the
value of $\hat{H}^{i=2}_\mathrm{max}$ just derived for the
secondary star. Specifically, we set
\begin{eqnarray}
\hat{H}_\mathrm{max}^{i=1} &=& \hat{H}^{i=2}_\mathrm{max} \biggl[
\frac{(x_\mathrm{max}^{i=1})^2 + 1}{(x_\mathrm{max}^{i=2})^2 + 1}
\biggr]^{1/2} \, . \label{eq:Hmax1}
\end{eqnarray}
Then we determine the value of the Bernoulli constant inside the primary star by examining the distribution of the variable $\hat{F}$ as defined in Eq.(48) of  \cite{hachisu86a}, that is,
\begin{eqnarray}
\hat{F}^{i} &\equiv& - \hat{\Phi} + \frac{1}{2}\hat{\Omega}^2
\hat{\varpi}^2 = (\hat{H}^{i} - \hat{C}^{i}) \, .  \label{eq:F}
\end{eqnarray}
In the vicinity of the original center of the primary star, that is,
in the vicinity of point $O_1$ as illustrated in Figure
\ref{fig:diagram}, the function $\hat{F}$ should exhibit a local
maximum.  We associate the location of this local maximum with the
updated position of point $O_1$ and we set
$\hat{F}^{i=1}_\mathrm{max}$ equal to the value of the function at
this local maximum. We therefore deduce from Eq.~(\ref{eq:F}) that,
\begin{eqnarray}
\hat{C}^{i=1} &=& \hat{H}^{i=1}_\mathrm{max} -
\hat{F}^{i=1}_\mathrm{max} .
\end{eqnarray}
With this constant in hand, the normalized enthalpy throughout the
primary star can be determined via the expression,
\begin{eqnarray}
\hat{H}^{i=1} &=& \hat{C}^{i=1} + \hat{F}^{i=1} \, ,
\end{eqnarray}
and we obtain an updated
``guess'' for the normalized density distribution inside the primary
star via the expression,
\begin{eqnarray}
\hat{\rho}^{i=1} &=& \frac{1}{ (x^{i=2}_{max})^3} \biggl\{ \biggl(
\frac{\hat{H}^{i=1}}{\hat{H}^{i=1}_\mathrm{max}} \biggr)^2
[(x^{i=1}_\mathrm{max})^2+1] -1 \biggr\}^{3/2} \, .\label{eq:rho1}
\end{eqnarray}

\subsection{Global Properties and Convergence}

Our iterative scheme is judged to be operating well if various
calculated model parameters --- such as the dimensionless
stellar masses $\hat{M}^i$ and Bernoulli constants $\hat{C}^i$
--- converge toward well-defined values. We also have found it
useful to track the convergence of various global energy
parameters.  Specifically, at the end of each iteration cycle
we calculate the dimensionless rotational kinetic energy
$\hat{K}$, gravitational potential energy $\hat{W}$, total
internal energy $\hat{U}$ (see, for example, Eq.~(75') in
Chapter $XI$ of \cite{Chandrasekhar67}), and globally averaged
pressure $\hat\Pi$ of the model, defined as follows:
\begin{eqnarray}
\hat{K} &\equiv& \int{\frac{1}{2} \hat{\Omega}^2 \hat{\varpi}^2
\hat\rho d\hat{V}} \, , \\
\hat{W} &\equiv& \int{\frac{1}{2} \hat\phi \hat\rho d\hat{V}} \, , \\
\hat{U} &\equiv& \int{\biggl[\biggl(\hat{H} - \frac{8A}{B}\biggr)\hat\rho -
\hat{P}\biggl]d\hat{V}} \, , \\
\hat\Pi &\equiv& \int{\hat{P}  d\hat{V}} \, .
\end{eqnarray}
where $d\hat{V} = \hat\varpi d\hat\varpi d\theta d\hat{z}$ is
the dimensionless differential volume element on our
cylindrical grid. Then the system's dimensionless total energy
is given by the sum,
\begin{eqnarray}
\hat{E}_\mathrm{tot} \equiv \frac{E_\mathrm{tot}}{G(\rho^{i=2}_{max})^2 \varpi_*^5}
&=& \hat{K} + \hat{W} + \hat{U} \, , \label{eq:Etot}
\end{eqnarray}
and, if the model has converged to a proper equilibrium state,
according to the virial theorem we should expect,
\begin{eqnarray}
2\hat{K} + \hat{W} + 3\hat\Pi &=& 0 \, . \label{virial}
\end{eqnarray}
In general, at each iteration step the condition of virial
equilibrium, Eq.~(\ref{virial}),  will not be satisfied, but if
our iteration scheme is well behaved, convergence toward the
virial condition should be achieved.  With this in mind, we
have found that the virial error,
\begin{eqnarray}
VE &\equiv& \biggl| \frac{2\hat{K} + \hat{W} + 3\hat\Pi}{\hat{W}} \biggr| \, ,
\label{virialError}
\end{eqnarray}
provides a meaningful measure of the quality of each model.

We declare that satisfactory convergence to a given model has
been achieved when the absolute value of the fractional change
between iterations has dropped below a specified convergence
criterion, $\delta \sim 10^{-4}$, for all of the following
quantities: $\hat{C}^i$, $\hat{M}^i$, $\hat\Omega$, $\hat{K}$,
$\hat{W}$, $\hat\Pi$, and the physical value of
$\varpi_\alpha$. In addition, the converged model is judged to
be a good equilibrium state if the virial error, VE, is
sufficiently small. Table \ref{tab:scf_convg9_6} illustrates
how we were able to achieve a lower virial error and, hence, a
more accurate representation of an equilibrium configuration,
by improving the grid resolution and/or by specifying a tighter
convergence criterion.  Specifically, the table shows that as
we were constructing binary model B3 (see discussion associated
with Table \ref{tab:scf_models_q_0.66}, below) we were able to
push the VE down from a value $\sim 5\times 10^{-3}$ to a value
$\sim 5\times 10^{-4}$ by increasing the grid resolution from
(64,128,33) to (128,256,65) zones in
($\hat\varpi$,$\theta$,$\hat{z}$) and by pushing $\delta$ from
$10^{-2}$ to $3.5\times 10^{-5}$.

After the SCF code has converged to the desired equilibrium
model, the various dimensionless variables are converted back
to proper physical units following, for example, the scalings
presented in Eqs.(22)-(27) of \cite{hachisu86a}.  We note in particular that the value of
the scale length $\varpi_* = \varpi_\alpha$ is obtained by
evaluating a dimensionless version of Eq.~(\ref{eq:enthalpy01}) for the secondary star in combination with Eq.(27) from \cite{hachisu86a}, which gives,
\begin{eqnarray}
\varpi_* = \biggl[ \frac{8A/B}{G\rho^{i=2}_{max} } \biggr]^{1/2}
( \hat{H}^{i=2}_{max} )^{-1/2} \biggl[(x^{i=2}_{max})^2 + 1
\biggr]^{1/4} \, .
\end{eqnarray}
In addition to the physical variables already identified, for
each converged model we have found it useful to evaluate the
system's total angular momentum,
\begin{eqnarray}
J_\mathrm{tot} &\equiv& \int{\varpi^2 \Omega \rho d{V}} \, ,
\end{eqnarray}
as well as the spin angular momentum of each component star,
$J_\mathrm{spin}^{i}$, and each star's Roche-lobe filling
factor, $f^i_\mathrm{RL}$.  As with the determination of
quantities such as $M^i$ and $R^i$, these latter two quantities
are obtained by performing volume integrals over appropriate
sub-domains of the computational grid, determined as follows.
Let the origin of a Cartesian grid coincide with the center of
mass of the binary system and align the $x$-axis of that grid
with the line that connects the centers of the two stars as
illustrated in Figure \ref{fig:diagram}. Between points $O_1$
and $O_2$ along this axis, the effective potential,
\begin{eqnarray}
\Phi_\mathrm{eff}(x) &\equiv& \Phi(x) - \frac{1}{2}\Omega^2 x^2 \, ,
\end{eqnarray}
will exhibit a maximum at position $x_{L1}$ associated with the
inner ``$L1$'' Lagrange point.  We define sub-domain
$\mathcal{D}_*^{i=2}$ as the volume of the grid for which $x
\equiv \varpi\cos\theta \geq x_{L1}$ and $\rho > 0$, that is,
the region occupied by the secondary star; and we define
sub-domain $\mathcal{D}_*^{i=1}$ as the volume of the grid for
which $x < x_{L1}$ and $\rho > 0$, that is, the region occupied
by the primary star. Then the mass of each star is determined
by the integral,
\begin{eqnarray}
M^i &=& \int_{\mathcal{D}_*^i}
{ \rho^i d{V}} \, ,
\end{eqnarray}
the volume occupied by each star is,
\begin{eqnarray}
\mathcal{V}^i_* &=& \int_{\mathcal{D}_*^i}
{ d{V}} \, ,
\end{eqnarray}
and the spin angular momentum of each star is given by the
expression,
\begin{eqnarray}
J_\mathrm{spin}^i &\equiv& \int_{\mathcal{D}_*^i}
{[\varpi^2 \sin^2\theta + (\varpi\cos\theta - \varpi_i)^2 ]
~\Omega \rho^{i} d{V}} \, .
\end{eqnarray}
Having determined the volumes $\mathcal{V}^i$ occupied by both
rotationally flattened and tidally distorted stars, we define
the mean radius of each star as,
\begin{eqnarray}
R^i &=& \biggl( \frac{3\mathcal{V}^i_*}{4\pi} \biggr)^{1/3} \, .
\end{eqnarray}

We furthermore define sub-domain
$\mathcal{D}^{i=2}_\mathrm{RL}$ as the volume of the grid for
which $x \geq x_{L1}$ and $\Phi_\mathrm{eff} \leq
\Phi_\mathrm{eff}(x_{L1})$, and sub-domain
$\mathcal{D}^{i=1}_\mathrm{RL}$ as the volume of the grid for
which $x < x_{L1}$ and $\Phi_\mathrm{eff} \leq
\Phi_\mathrm{eff}(x_{L1})$.  Then the Roche-lobe volume
surrounding each star is,
\begin{eqnarray}
\mathcal{V}^i_\mathrm{RL} &=& \int_{\mathcal{D}_\mathrm{RL}^i}
{ d{V}} \, ,
\end{eqnarray}
and each star's Roche-lobe filling factor is obtained from the
ratio,
\begin{eqnarray}
f^i_\mathrm{RL} &=& \frac{\mathcal{V}_*^i}{\mathcal{V}_\mathrm{RL}^i} \, .
\end{eqnarray}

\section{Results}
\subsection{Single White Dwarfs}\label{sec:SingleWDs}
\begin{figure}[ht]
\epsfig{file=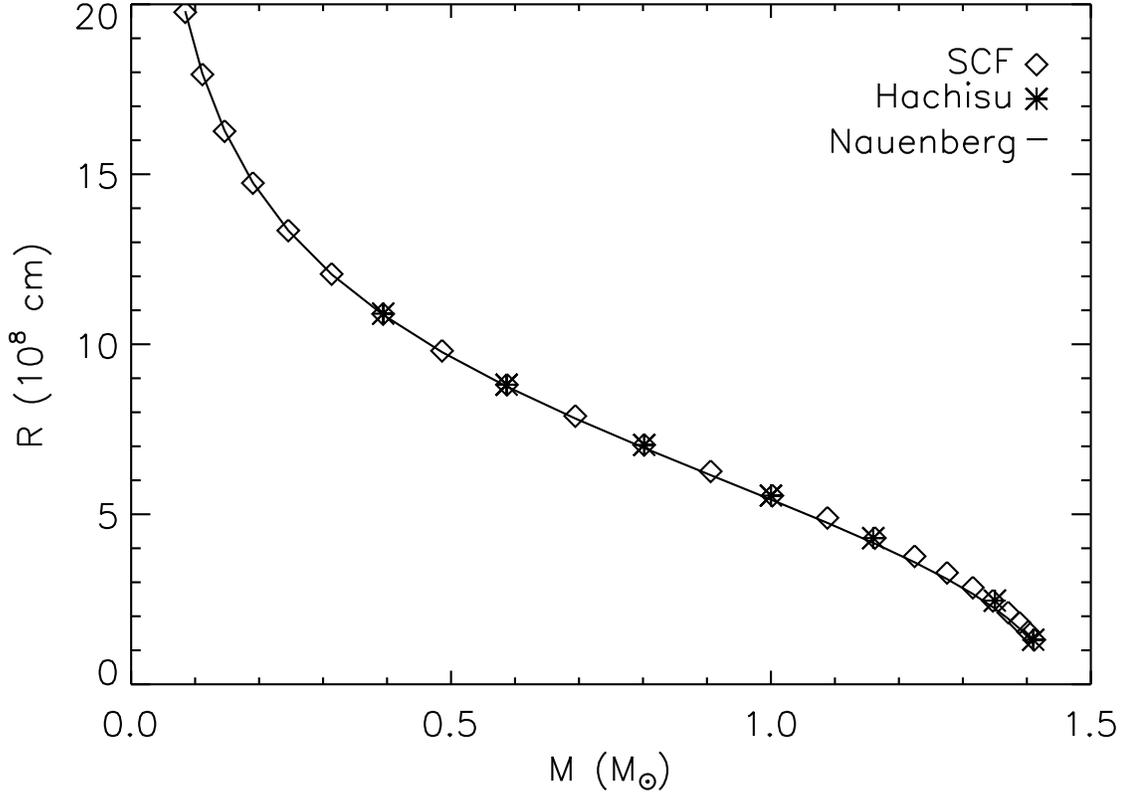,width=\textwidth}
\figcaption[f1.ps]{The mass-radius relationship is
shown for spherical stars with our adopted ZTWD equation of
state. Diamonds represent results derived using our
three-dimensional SCF scheme applied to nonrotating, isolated
configurations (see Table \ref{tab:scf_single}); asterisks show
previously published results for the same equation of state
taken from \cite{hachisu86a}; the solid curve shows the
analytic mass-radius relation,
Eq.~(\ref{eq:MRnauenbergAppendix}), derived by
\cite{Nauenberg72}. \label{fig:MR}}
\end{figure}

\begin{figure}[ht]
\epsfig{file=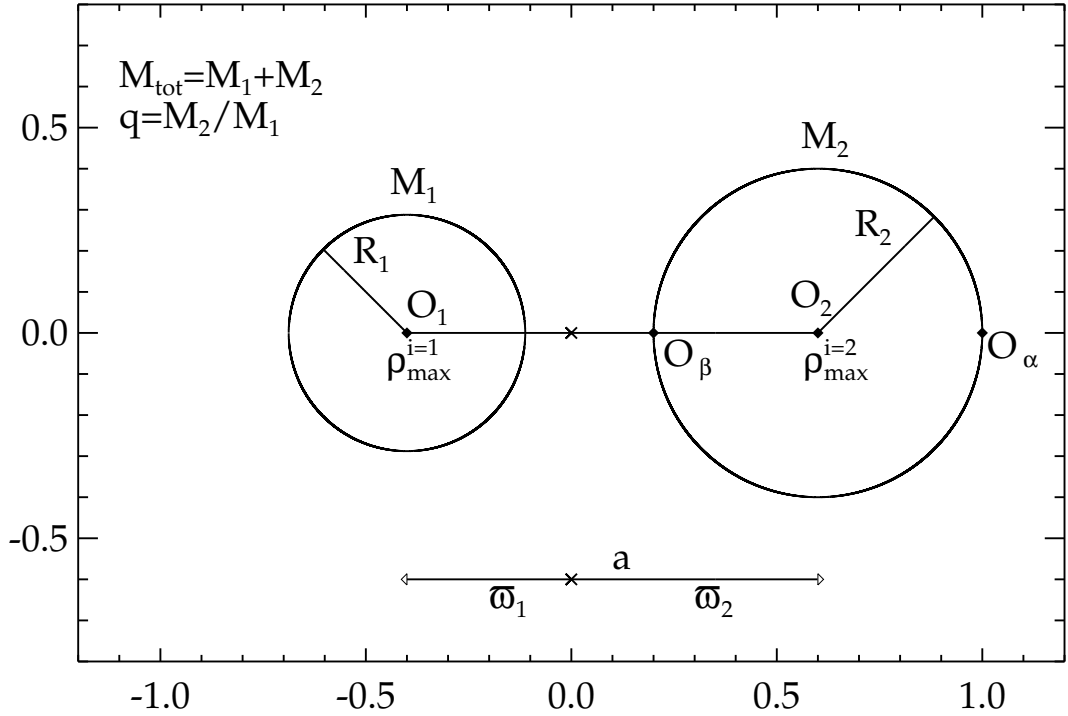,width=\textwidth}
\figcaption[f2.ps]{Schematic diagram illustrating the equatorial-plane
structure of a binary star system.  The primary star, on the left, has a mass
$M_1$, a radius $R_1$, and a central density $\rho_\mathrm{max}^{i=1}$; the
secondary star, on the right, has a mass $M_2 \leq M_1$,
a radius $R_2$, and a central density $\rho_\mathrm{max}^{i=2}$.
The centers of mass of the two stars (points labeled $O_1$ and $O_2$) are
separated by a distance $a = \varpi_1+\varpi_2$, and their distances from
the center of mass of the system are, respectively, $\varpi_1$ and
$\varpi_2$.  The points labeled {\bf $O_\alpha$} and {\bf $O_\beta$}
identify, respectively, the outer edge and inner edge of the secondary star.
\label{fig:diagram}}
\end{figure}

As mentioned earlier, we initially used a simplified version of
our SCF code to construct a large number of single, nonrotating
white dwarfs in order to compare our solutions with previous
results (see Figure \ref{fig:MR}) and to provide initial
guesses for the density distributions inside both stars in each
binary system. Table \ref{tab:scf_single} details the
properties of single, nonrotating white dwarfs that have
central densities ranging from $10^{4.5}~\mathrm{g}
~\mathrm{cm}^{-3}$ to $10^{10} ~\mathrm{g}~\mathrm{cm}^{-3}$ as
determined from our model calculations; the 23 selected models
are equally spaced in units of $\log{\rho_\mathrm{max}}$. These
spherical models were constructed on a uniform cylindrical mesh
with resolution $(128, 128, 128)$ in
$(\hat\varpi,\theta,\hat{z})$ using a convergence criterion
$\delta = 10^{-4}$. For each converged model, the first six
columns of Table \ref{tab:scf_single} list, respectively, the
star's mass $M$ in solar masses, radius $R$ in units of $10^8$
cm, central density $\rho_\mathrm{max}$ in
$\mathrm{g}~\mathrm{cm}^{-3}$, corresponding value of
$x_\mathrm{max} = (\rho_\mathrm{max}/B)^{1/3}$, moment of
inertia,
\begin{eqnarray}
I = \int{\varpi^2 \rho dV} \, ,
\end{eqnarray}
in units of $10^{50}~\mathrm{g}~\mathrm{cm}^2$, and the radius
of gyration, $k \equiv I/(MR^2)$. As shown in the last column
of Table \ref{tab:scf_single}, a typical virial error for these
converged models was $10^{-4} - 10^{-5}$. 
The values tabulated for the radius of gyration vary smoothly
from $k = 0.2036$ for $M = 0.0844 M_\odot$ to $k = 0.1013$ for
$M = 1.4081 M_\odot$. This is consistent with our understanding
that low-mass white dwarfs have structures similar to $n=3/2$
polytropes for which $k = 0.205$ \citep{Rucinski88}, while
high-mass white dwarfs display structures similar to $n=3$
polytropes for which $k = 0.0758$ \citep{Rucinski88}. Our
values of $k$ over this range of stellar masses are also
consistent with the analytic function for $k(M)$ that
\cite{MNS04} fit through similar spherical model data.
Knowledge of the radius of gyration of these spherical ZTWD
models has assisted us in analyzing the tidally distorted
structures that arise in our models of synchronously rotating
white dwarfs in close binary systems (see further discussion,
below).

Using this same three-dimensional, cylindrical coordinate grid
we constructed nonrotating models with central densities above
$10^{10}~\mathrm{g}~\mathrm{cm^{-3}}$, that is, with masses
above $1.4 M_\odot$.  We have not included these higher mass
models in Table \ref{tab:scf_single} or Figure \ref{fig:MR},
however, because they did not converge to satisfactorily
accurate structures. In particular, as the mass was steadily
increased above $1.4 M_\odot$, the models converged to
structures with steadily increasing (rather than decreasing)
values of $k$.  By contrast, models constructed using a
one-dimensional spherical code with much higher spatial
resolution displayed values of $k$ that decreased steadily to a
value of 0.0755 at masses approaching $M_\mathrm{ch}$. If
desired, the three-dimensional computational grid resolution
could be increased to produce more accurate models of the white
dwarf structure near the Chandrasekhar mass limit.

\subsection{White Dwarf Binary Sequences}

The slow inspiral evolution of a DWD binary can be mimicked by
constructing a sequence of detached binaries having fixed
$M_\mathrm{tot}$ and fixed $q$  but varying separation, down to
the separation at which the less massive star first makes
contact with its Roche lobe.  In an effort to illustrate the
capabilities of our code, we have constructed three binary
sequences having the same total mass --- namely,
$M_\mathrm{tot} = 1.5 M_\odot$ --- but three separate mass
ratios. Specifically, sequence `A' has $q=1$, sequence `B' has
$q = 2/3$, and sequence `C' has $q=1/2$. As detailed in Table
\ref{tab:scf_single2}, spherical models were constructed with
the desired primary and secondary masses for these three
sequences --- specifically, $M = 0.5 M_\odot, 0.6 M_\odot, 0.75
M_\odot, 0.9 M_\odot$ and $1.0 M_\odot$ --- to provide good
``guesses'' for the initial binary star density distributions
to start each SCF iteration.  In addition to listing the values
of $M$, $R$, $\rho_\mathrm{max}$, $x_\mathrm{max}$, and $k$ for
each of these converged spherical models, as was done for a
wider range of spherical models in Table \ref{tab:scf_single},
Table \ref{tab:scf_single2} also lists values for the global
energies $W$, $U$, and $\Pi$ in units of $10^{50}$ ergs.

Along each sequence, all the binary models were constructed
using a uniform cylindrical grid with (128,256,65) zones in
$(\hat\varpi,\theta,\hat{z})$; by implementing reflection
symmetry through the equatorial plane, only half as many zones
were needed in the vertical direction as in the radial
direction to achieve the same resolution in both.  No additional symmetries were assumed in constructing the sequence, although, for
the models shown here, symmetry through the x-z plane also could have been implemented for additional savings.  The
convergence criterion was set to $\delta = 2.5 \times 10^{-4}$;
in most models, $\hat\Omega$ was the last variable to converge
to this desired level. We note that, because the same number of
grid zones was used for each model and each binary was scaled
to fit entirely within the grid, the effective resolution of
each star decreased as the binary separation $a$ increased
along each sequence.

Two tables have been produced for each DWD inspiral sequence in
order to detail the properties of the models that lie along
each sequence.  For sequence `A' ($q = 1$), for example, Table
\ref{tab:scf_models_q_1.00} lists the values of six global
binary system parameters ($a$, $\Omega$, $M_\mathrm{tot}$, $q$,
$J_\mathrm{tot}$, $E_\mathrm{tot}$) and the virial error
obtained for thirty-five models (numbered $A1$ through $A35$)
whose binary separations vary from $2.0956\times
10^9~\mathrm{cm}$ at contact (model $A1$) to $3.0911\times
10^9~\mathrm{cm}$ (model $A35$). For this same group of models,
Table \ref{tab:scf_models_ind_q_1.00} lists calculated values
of five parameters ($M_i$, $R_i$, $\rho_\mathrm{max}^{i}$,
$f_\mathrm{RL}^{i}$, $J_\mathrm{spin}^i$) for the individual
stellar components ($i=1,2$).  Tables
\ref{tab:scf_models_q_0.66} and \ref{tab:scf_models_ind_q_0.66}
provide the same detailed information for models along sequence
`B' ($q=2/3$), and Tables \ref{tab:scf_models_q_0.50} and
\ref{tab:scf_models_ind_q_0.50} provide this information for
models along sequence `C' ($q = 1/2$).

The equatorial-plane density distributions displayed in Figures
\ref{fig:DenSeries100}, \ref{fig:DenSeries066}, and
\ref{fig:DenSeries050} illustrate the degree to which both
white dwarf components are distorted by tides for various
binary separations along each sequence.  Labels in the
upper-right-hand corner of each figure panel identify each
binary system according to its corresponding position along
each sequence, as itemized in Tables
\ref{tab:scf_models_q_1.00} - \ref{tab:scf_models_ind_q_0.50}.
Along sequence `A' (Figure \ref{fig:DenSeries100}), both
components of the binary system are of equal size and display
identical degrees of tidal distortion because the mass ratio $q
= 1$.  Along sequences `B' and `C' (Figures
\ref{fig:DenSeries066} and \ref{fig:DenSeries050},
respectively), however, the primary star (on the left in each
figure panel) is noticeably smaller and less distorted than the
secondary star.

Figure \ref{fig:075_075series} has been constructed from the
data detailed in Tables \ref{tab:scf_models_q_1.00} and
\ref{tab:scf_models_ind_q_1.00} for binary sequence `A.'
Specifically, the diamond symbols in the top two panels and in
the bottom panel of this figure show, respectively, how the
binary system's total angular momentum, $J_\mathrm{tot}$, total
energy, $E_\mathrm{tot}$, and orbital angular velocity,
$\Omega$, vary with orbital separation along this sequence; and
the third panel from the top shows how the Roche-lobe filling
factor, $f_\mathrm{RL}^i$, varies with orbital separation for
both the primary star (diamonds) and the secondary star
(asterisks). Figures \ref{fig:090_060series} and
\ref{fig:100_050series} have been similarly constructed from
the data detailed, respectively, in Tables
\ref{tab:scf_models_q_0.66} and
\ref{tab:scf_models_ind_q_0.66}, and in Tables
\ref{tab:scf_models_q_0.50} and
\ref{tab:scf_models_ind_q_0.50}.

Following the lead of \cite{New97}, in constructing Figures
\ref{fig:075_075series} - \ref{fig:100_050series} we have
normalized our tabulated values of $J_\mathrm{tot}$ and
$E_\mathrm{tot}$ to the quantities,
\begin{eqnarray}
J_\mathrm{norm} &\equiv& (G M_{0.75}^3 R_{0.75})^{1/2} = 4.0735 \times
10^{50} ~\mathrm{g}~\mathrm{cm}^2~\mathrm{s}^{-1} \, , \\
E_\mathrm{norm} &\equiv& \frac{GM_{0.75}^2}{R_{0.75}} = 2.0119 \times
10^{50} ~\mathrm{erg} \, ,
\end{eqnarray}
where
$R_{0.75} = 7.4244\times 10^{8}$ cm is the radius of a
spherical ZTWD whose mass is $M_{0.75} = 0.7522~M_\odot$ as
tabulated in Table \ref{tab:scf_single2}. Also, at each
separation our tabulated values of $\Omega$ have been
normalized to the Keplerian orbital frequency,
\begin{eqnarray}
\Omega_\mathrm{K} &=& \biggl( \frac{2G M_{0.75}}{a^3}
\biggr)^{1/2} \, .
\end{eqnarray}
In all three figures, values of the orbital separation have
been specified (bottom horizontal axis) in units of
$10^9~\mathrm{cm}$ and (top horizontal axis) as normalized to
the radius of a spherical ZTWD having the mass of the system's
secondary star as tabulated in Table \ref{tab:scf_single2},
that is, $R_{0.75} \equiv 7.424 \times 10^{8}~\mathrm{cm}$,
$R_{0.60}= 8.671\times 10^{8}~\mathrm{cm}$ and $R_{0.50}=
9.638\times 10^{8}~\mathrm{cm}$.

\cite{New97} have previously constructed inspiral sequences for
{\it equal-mass} DWD binary systems in which the structure of
the individual component stars is governed by the Chandrasekhar
ZTWD equation of state (\ref{eq:EOSchandra}). The sequences
published by \cite{New97} cover a wide range of total masses.
The one that most closely resembles our sequence `A' (our only
equal-mass sequence) has $M_\mathrm{tot} = 1.63 M_\odot$; the
functional behavior of $E_\mathrm{tot}(a)$ and
$J_\mathrm{tot}(a)$ for this sequence is presented in Figure 16
of \cite{New97}.  Along this $M_\mathrm{tot}=1.63 M_\odot$
sequence, the two stars first make contact with their
respective Roche lobes at a normalized separation of
approximately 2.825 (see also Figure 5 of New \& Tohline 1997).
This is completely consistent with the behavior of our sequence
`A,' where contact occurs (model A1) when $a/R_{0.75} = 2.823$.

The DWD sequences constructed by \cite{New97} all extend to
separations smaller than the point of first contact, as their
SCF technique allowed them to build over-contact (common
envelope) equal-mass binaries.  Their functions
$E_\mathrm{tot}(a)$ and $J_\mathrm{tot}(a)$ display a quadratic
behavior along the over-contact segment of each sequence,
passing through a minimum at a binary separation smaller than
the point of first contact. None of our three sequences show
this behavior because we have not attempted to construct models
past the initial point of contact. Indeed, it seems unlikely
that equilibrium configurations exist at smaller separations
except when the system mass ratio is precisely $q=1$.

For each of our DWD binary sequences, it is useful to compare
the displayed functional behavior of $J_\mathrm{tot}(a)$ from
our numerical models against the behavior predicted by two
simplified models. In the case of two point masses in circular
orbit, the total angular momentum $J_\mathrm{pm}$ is given
simply by the system's orbital angular momentum, that is,
\begin{eqnarray}
  J_\mathrm{pm} = J_\mathrm{orb} &=& M_1 \varpi_1^2 \Omega_\mathrm{K} +
  M_2 \varpi_2^2 \Omega_\mathrm{K} \nonumber \\
  &=& \frac{q}{(1+q)^2}  \biggl[ GM_\mathrm{tot}^3 a
  \biggr]^{1/2} \label{eq:Jpm}
\end{eqnarray}
This function, normalized to $J_\mathrm{norm}$, is displayed by
the solid curve in the top panels of Figures
\ref{fig:075_075series} - \ref{fig:100_050series}.  An even
more realistic representation of the function
$J_\mathrm{tot}(a)$ can be obtained by adding an approximate
representation for the spin angular momentum, $I_i\Omega$, of
both stars to the point-mass expression for $J_\mathrm{orb}$.
If we assume that both stars retain a spherical structure while
spinning at the Keplerian orbital frequency,
$\Omega_\mathrm{K}$, the appropriate expression for the total
``spinning sphere'' system angular momentum is,
\begin{eqnarray}
  J_{ss} &=&  J_\mathrm{orb} + (I_1+I_2)\Omega_\mathrm{K} =
  J_\mathrm{orb} + \biggl( k_1 M_1 R_1^2 + k_2 M_2 R_2^2 \biggr)
  \Omega_\mathrm{K}  \nonumber  \\
  &=& J_\mathrm{pm} \biggl\{ 1 + \frac{(1+q)}{q} \biggl[ k_1
  \biggl(\frac{R_1}{a}\biggr)^2 + q k_2  \biggl(\frac{R_2}{a}\biggr)^2
  \biggr] \biggr\}  \label{eq:Jss}  \, ,
\end{eqnarray}
where, in addition to $q$, values of (the constants) $R_i$ and
$k_i$ appropriate for each binary sequence can be obtained from
Table \ref{tab:scf_single2}. Function (\ref{eq:Jss}),
normalized to $J_\mathrm{norm}$, is displayed by the dot-dashed
curve in the top panels of Figures \ref{fig:075_075series} -
\ref{fig:100_050series}. Analytic expression (\ref{eq:Jpm})
predicts that $J_\mathrm{tot} \propto a^{1/2}$.  Through a
correction factor, Eq.~(\ref{eq:Jss}) displays a somewhat more
complex behavior. Overall, our SCF model sequences match
Eq.~(\ref{eq:Jss}) particularly well. The largest deviation
arises in all cases at the smallest separations; the slope of
the SCF-generated $J_\mathrm{tot}(a)$ function flattens
somewhat as the secondary star approaches contact with its
Roche lobe, that is, as $f_\mathrm{RL}^{i=2}\rightarrow 1$.

The functional dependence of each system's total energy,
$E_\mathrm{tot}(a)$, can be understood in a similar fashion.
Considering only the kinetic and gravitational potential energy
of two point masses in circular orbit, we obtain,
\begin{eqnarray}
E_\mathrm{orb} &=& K_\mathrm{orb} + W_\mathrm{orb} = - K_\mathrm{orb}
= - \frac{1}{2} \biggl[ \frac{q}{(1+q)^2} \biggr]
\frac{G M_\mathrm{tot}^2}{a} \, ,\label{eq:Eorb}
\end{eqnarray}
where we have used the virial relation $(2K_\mathrm{orb} +
W_\mathrm{orb}) = 0$.  While this $a^{-1}$ functional
dependence explains the general $E_\mathrm{tot}(a)$ behavior
exhibited in Figures \ref{fig:075_075series} -
\ref{fig:100_050series} by our numerically constructed model
sequences, expression (\ref{eq:Eorb}) is missing a nontrivial
shift in the overall energy scale that is set by the binding
energies of the two stars, namely,
\begin{eqnarray}
E_\mathrm{b} &=&  \sum_{i=1}^2 \biggl( W^i + U^i \biggr) \, . \label{eq:Eb}
\end{eqnarray}
Based on the properties of the spherical stellar models
provided in Table \ref{tab:scf_single2}, the appropriate energy
shift for sequences `A,' `B,' and `C' is, respectively, $E_b =
-1.551 \times 10^{50}~\mathrm{ergs}$, $-1.698 \times
10^{50}~\mathrm{ergs}$, and $-1.963 \times
10^{50}~\mathrm{ergs}$.  Adding $E_\mathrm{b}$ to
$E_\mathrm{orb}$ provides what we will refer to as the ``point
mass'' total system energy,
\begin{eqnarray}
E_\mathrm{pm} &=& - \frac{1}{2} \biggl[ \frac{q}{(1+q)^2} \biggr]
\frac{G M_\mathrm{tot}^2}{a} + E_\mathrm{b} \, .\label{eq:Epm}
\end{eqnarray}
This analytic function, normalized to $E_\mathrm{norm}$, is
displayed as a solid curve in the plots of $E_\mathrm{tot}$
versus $a$ shown in Figures \ref{fig:075_075series} -
\ref{fig:100_050series}. An improved approximation that we will
refer to as the ``spinning sphere'' total system energy can be
obtained by adding the rotational kinetic energy of both stars,
assuming they remain spherically symmetric and spin uniformly
with the Keplerian orbital frequency.  Specifically,
\begin{eqnarray}
E_\mathrm{ss} &=& E_\mathrm{pm} + \sum_{i=1}^{2} \biggl( \frac{1}{2}I_i
\Omega_\mathrm{K}^2  \biggr) \nonumber \\
&=&  E_\mathrm{b} + E_\mathrm{orb} \biggl\{ 1 - \frac{(1+q)}{q}
\biggl[ k_1 \biggl( \frac{R_1}{a} \biggr)^2 +
q k_2 \biggl( \frac{R_2}{a} \biggr)^2 \biggr]
\biggr\} \, . \label{eq:Ess}
\end{eqnarray}
This function, normalized to $E_\mathrm{norm}$, is displayed as
a dot-dashed curve in the plots of $E_\mathrm{tot}$ versus $a$
shown in Figures \ref{fig:075_075series} -
\ref{fig:100_050series}.  Expression (\ref{eq:Ess}) describes
particularly well the variation of $E_\mathrm{tot}$ with
separation displayed by our numerically constructed binary
sequences `B' and `C.'  We note, however, that all three of our
sequences show that the total system energy drops slightly
below the behavior predicted by Eq.~(\ref{eq:Ess}) at the
smallest separations.

The curve outlined by asterisks in the third panel from the top
of Figures \ref{fig:075_075series} - \ref{fig:100_050series}
shows that $f_\mathrm{RL}^{i=2}$ steadily increases from a
value $\sim 0.2$ to a value of $1.0$ at the smallest separation
along all three inspiral sequences, implying that the secondary
star has made contact with its Roche lobe.  For comparison, the
curve outlined by diamonds in the same panel of these three
figures shows how the Roche-lobe filling factor of the primary
star varies along each sequence.  The value of
$f_\mathrm{RL}^{i=1}$ does not climb above 0.063 for sequence
`C' or above 0.191 for sequence `B,' reflecting the fact that
in both cases the primary star is significantly more massive
--- and, hence, it has a significantly smaller radius --- than
the secondary star.  For inspiral sequence `A,'
$f_\mathrm{RL}^{i=1}(a)$ displays an identical behavior to
$f_\mathrm{RL}^{i=2}(a)$ because the primary and secondary
stars have equal masses.

The bottom panel of Figures \ref{fig:075_075series} -
\ref{fig:100_050series} displays the behavior of the normalized
orbital frequency $\Omega/\Omega_\mathrm{K}$ as a function of
binary separation derived from our three numerically
constructed inspiral sequences.  At the smallest separations,
our models show that the orbital frequency is always $\sim
0.5\%$ higher than predicted by the ``point-mass'' Keplerian
frequency.  Our equal-mass sequence exhibits the largest
deviation at contact; specifically, for model `A1,' we find
$\Omega = 1.0085 ~\Omega_\mathrm{K}$.  As the separation is increased along
each sequence, the figures show that $\Omega/\Omega_\mathrm{K}$
approaches unity, as expected. However, at a sufficiently wide
separation, each of our sequences displays a tiny discontinuous
drop in the orbital frequency, followed by further decline that
ultimately falls below the local Keplerian value.
 We suspect
this odd behavior at wide separations arises from the discrete
nature of our grid calculations coupled with progressively
fewer grid zones falling inside both stars --- resulting in
progressively poorer numerical resolution --- at wider
separations.

\subsection{Conservative Mass-Transfer Sequences}

During a phase of stable mass transfer, a DWD binary system
will evolve in such a way that the secondary star remains in
marginal contact with its Roche lobe while it slowly transfers
mass to the primary star.  If the total mass of the system is
conserved, then the evolution should proceed along a sequence
of synchronously rotating configurations in which
$M_\mathrm{tot}$ is constant, $f_\mathrm{RL}^{i=2} = 1$, and
$q$ is steadily decreasing. Models $A1$, $B1$ and $C1$ can be
viewed as representing three such configurations along a
sequence whose total system mass is $M_\mathrm{tot}=1.5
M_\odot$.  In evolving from an initially equal-mass, contact
configuration (model $A1$) to a semi-detached configuration
with $q=2/3$ (model $B1$), then on to a semi-detached
configuration with $q=1/2$ (model $C1$), the separation of such
a system (measured in units of $10^9~\mathrm{cm}$) will
increase from $a_9 = 2.10$, to $a_9 = 2.67$, then to $a_9 =
3.18$; and the system's orbital period ($P_\mathrm{orb} =
2\pi/\Omega$) will increase from 42.3 s to 61.0 s, then to 79.5
s.

It is clear, therefore, that our new SCF code can be used to
construct model sequences that mimic the evolution of DWD
systems undergoing slow, conservative mass-transfer.  The
models detailed in Tables \ref{tab:contactD} (sequence `D') and
\ref{tab:contactE} (sequence `E') trace two such semi-detached
sequences as the system mass ratio evolves from $q=1$ to $q
\lesssim 0.5$. For sequence `D,' $M_\mathrm{tot} = 1.5 M_\odot$
and for sequence `E,' $M_\mathrm{tot}=1.0 M_\odot$. In the top
two panels of Figures \ref{fig:contactM1.5} and
\ref{fig:contactM1.0}, data from Tables \ref{tab:contactD} and
\ref{tab:contactE} have been plotted as diamond symbols to
illustrate how $a$ and $\Omega$ vary with $q$ while
$f_\mathrm{RL}^{i=2}$ is held to a value of unity (definition
of a semi-detached binary) along these two fixed-mass
sequences.

Up to now, the community has relied upon some relatively simple
analytic expressions to approximate the behavior of, for
example, $a(q)$ along conservative mass-transfer evolutionary
trajectories.  For example, by setting the radius of the
secondary star as given by the Nauenberg mass-radius relation
(\ref{eq:MRnauenbergAppendix}) equal to the Roche-lobe radius
$R_\mathrm{RL}$ as defined in terms of $a$ and $q$ by the
approximate relation provided by \cite{Eggleton83}, namely,
\begin{eqnarray}
R_\mathrm{RL} &=& a \biggl[ \frac{0.49 q^{2/3}}{0.6 q^{2/3}
+ \ln(1 + q^{1/3})} \biggr] \, ,
\end{eqnarray}
one obtains,
\begin{eqnarray}
\frac{a}{R_\odot} &\approx&  0.0229 (n_2q^2)^{-1/3} (1 - n_2^{4/3})^{1/2}
\biggl[ 0.6 q^{2/3} + \ln(1 + q^{1/3}) \biggr]   \, , \label{eq:aofqm}
\end{eqnarray}
where,
\begin{eqnarray}
n_2 &\equiv& \frac{q}{(1+q)} \biggl( \frac{M_\mathrm{tot}}{M_\mathrm{ch}}
\biggr) \, .
\end{eqnarray}
The function $a(q)$, defined by Eq.~(\ref{eq:aofqm}) for a
given $M_\mathrm{tot}$, has been plotted as a solid curve in
the top panels of Figures \ref{fig:contactM1.5} and
\ref{fig:contactM1.0}, and the Keplerian orbital frequency
associated with this separation (and relevant $M_\mathrm{tot}$)
has been plotted as a solid curve in the second panel of
Figures \ref{fig:contactM1.5} and \ref{fig:contactM1.0}.  For
both sequence `D' and sequence `E,' the analytically derived
curves are consistently offset by 3 - 5\% from our numerical
model results. But overall, the analytically predicted
functional behavior of $a(q)$ and $\Omega(q)$ is in very good
agreement with our results.  This is reassuring as it provides
a degree of validation for both our numerical code and the
approximations that were adopted by earlier investigators when
deriving the more easily manipulated analytic expressions.

Finally, in the bottom two panels of Figures
\ref{fig:contactM1.5} and \ref{fig:contactM1.0}, the diamond
symbols display the variation of $J_\mathrm{tot}$ and
$E_\mathrm{tot}$ with $q$ along sequence `D' and sequence `E,'
respectively. The solid curve drawn in the $J_\mathrm{tot}(q)$
panel of both figures shows the behavior predicted by our
``spinning sphere'' expression for the total system angular
momentum (\ref{eq:Jss}) when used in conjunction with the
$a(q,M_\mathrm{tot})$ behavior prescribed by
Eq.~(\ref{eq:aofqm}).  Again, for a given $M_\mathrm{tot}$
there appears to be very good agreement between the functional
behavior of $J_\mathrm{tot}(q)$ displayed by our numerical
model results and the analytic expressions.  There is also a
systematic offset between the two.  In either case it is clear
that, unlike the behavior displayed by $a(q)$ and $\Omega(q)$,
the system's total angular momentum does not vary monotonically
with $q$ along a conservative mass-transfer evolutionary
trajectory.  Note, in particular, that if the system mass ratio
$q$ is initially close to unity, $J_\mathrm{tot}$ {\it
increases} as $q$ decreases along the displayed trajectory.
This result is unphysical.  It signifies that slow evolution
along a synchronously rotating, conservative mass-transfer
trajectory can occur only if, at the onset of mass-transfer, $q
< q_\mathrm{crit}$, where the value of $q_\mathrm{crit}$ for a
given $M_\mathrm{tot}$ is prescribed by the location of the
maximum of the $J_\mathrm{tot}(q)$ curve.  For our model
sequences `D' and `E,' we see that $q_\mathrm{crit} \lesssim
2/3$, consistent with the mass-transfer stability limit that
has already received much attention in the literature
\citep{FTE09}.

\section{Summary and Conclusions}

Based on the earlier work of the Hachisu (1986a,b) and \cite{hachisuEtAl86a} we have developed a self-consistent-field technique that can be
used to construct equilibrium models of synchronously rotating
DWD binaries having a range of total masses, mass ratios, and
binary separations. In addition to effects introduced by
synchronous rotation, the distorted structure of both stars in
each converged model is governed by the zero-temperature white
dwarf equation of state (\ref{eq:EOSchandra}) and a
self-consistently determined, Newtonian gravitational field. In
an effort to illustrate the technique's capabilities, we have
constructed a set of models along five sequences: Three
sequences (`A', `B', and `C') mimic the last segment of the
detached ``inspiral'' phase of DWD binary evolutions during
which both $M_\mathrm{tot}$ and $q$ are held constant as $a$
decreases; and two sequences (`D' and `E') mimic a
semi-detached ``conservative mass transfer'' phase of evolution
during which $M_\mathrm{tot}$ is held fixed and the less
massive star stays in marginal contact with its Roche lobe, but
$q$ steadily decreases while $a$ steadily increases.

Along each inspiral sequence, the functional dependence of
$J_\mathrm{tot}$ and $E_\mathrm{tot}$ on the orbital separation
can be well understood in terms of simple analytical
expressions that describe two spinning spherical white dwarfs
in circular orbit about one another.  For a given total mass
and separation, the calculated orbital frequencies along each
inspiral sequence deviate measurably from associated Keplerian
frequencies only in models for which the Roche-lobe filling
factor of the less-massive star is $\gtrsim 60\%$.  But, at
least for the sequences examined here, the deviation from
Keplerian frequencies is never more than 1\% even at contact.

Along both conservative mass-transfer sequences, we have
documented how $a$, $\Omega$, $J_\mathrm{tot}$ and
$E_\mathrm{tot}$ vary with the system mass ratio as $q$
decreases by roughly a factor of two, from $q = 1.0$ down to $q
\lesssim 0.5$.  Along each sequence we have compared our
numerically determined values of $a$ at various values of $q$
with the analytic $a(q)$ function (\ref{eq:aofqm}) that is
derived by setting the radius of the less massive star, as
specified by the \cite{Nauenberg72} mass-radius relation, equal
to the Roche-lobe radius, as approximated by \cite{Eggleton83}.
Qualitatively, our results show the same $a(q)$ behavior that
is predicted by this analytic expression. However, at a given
$q$ the value of $a$ derived from our models is consistently
$\sim 8\%$ larger than the value obtained from
Eq.~(\ref{eq:aofqm}).  The analytic expression could be brought
into closer quantitative agreement with our numerical results
if the leading coefficient in Eq.~(\ref{eq:aofqm}) is
increased by $8\%$, that is, if the expression's leading
coefficient is changed from 0.0229 to 0.0247.  This
modification will, in turn, decrease the Keplerian frequency
obtained from the analytic $a(q)$ expression by $\sim 9\%$,
simultaneously bringing the analytically predicted orbital
frequency into much closer agreement with our numerically
determined values of $\Omega$.  Along both of our conservative
mass-transfer sequences, the plot of $J_\mathrm{tot}(q)$
displays an extremum at a value of $q \lesssim 2/3$.  The
location of this extremum is almost certainly identifying the
value of $q_\mathrm{crit}$ that is relevant along both
sequences.

The development of this SCF technique was originally motivated
by our desire to build models that would serve as good,
``quiet'' initial conditions for hydrodynamical simulations
that are designed to probe the onset and nonlinear development
of mass-transfer instabilities in close, unequal-mass DWD
binaries. The new computational tool that we have described in
this paper achieves this objective.

\acknowledgments We acknowledge valuable interactions that we
have had with B. Bourdin, J. Frank, D. Marcello, P. M. Motl, and
S. Ou over the course of this project.  We also thank an anonymous referee for pointing us to a key 
reference from Hachisu's collection of work during the mid-1980's.  This work has been
supported, in part, by grants AST-0708551 and DGE-0504507 from
the U.S. National Science Foundation and, in part, by grant
NNX07AG84G from NASA's ATP program.  This research also has
been made possible by grants of high-performance computing time
on the TeraGrid (MCA98N043), at LSU, and across LONI (Louisiana
Optical Network Initiative).

\appendix
\section{White Dwarf Mass-Radius
Relationship\label{Ap:MRrelationship}}

\subsection{The Chandrasekhar Mass}

\cite{Chandrasekhar35} was the first to construct models of
spherically symmetric stars using the equation of state defined
by Eq.~(\ref{eq:EOSchandra}) and, in so doing, demonstrated
that the maximum mass of an isolated, nonrotating white dwarf
is $M_\mathrm{ch} =1.44 (\mu_e/2) M_\odot$, where $\mu_e$ is
the number of nucleons per electron and, hence, depends on the
chemical composition of the WD. A concise derivation of
$M_\mathrm{ch}$ (although, at the time, it was referred to as
$M_3$) is presented in Chapter $XI$ of \cite{Chandrasekhar67},
where we also find that the expressions for the two key
coefficients in Eqs.~(\ref{eq:EOSchandra}) and
(\ref{eq:DefineX}) are,
\begin{eqnarray}
A &\equiv& \frac{\pi m_e^4 c^5}{3h^3}  \, , \\
B\mu_e^{-1} &\equiv& \frac{8\pi m_p }{3}\biggl( \frac{m_e c}{h} \biggr)^3
\, .
\end{eqnarray}
Numerical values for $A$ and $B\mu_e^{-1}$ are given here in
Table \ref{tab:physicalConstants} along with values of the
physical constants $c$, $h$, $m_e$, and $m_p$ that we have used
(column 2) and that \cite{Chandrasekhar67} used (column 3) to
determine the values of $A$ and $B\mu_e^{-1}$. The derived
analytic expression for the limiting mass is,
\begin{eqnarray}
\mu_e^2 M_\mathrm{ch} &=& 4\pi  m_3 \biggl( \frac{2A}{\pi G}
\biggr)^{3/2} \frac{\mu_e^2}{B^2} = 1.14205 \times 10^{34} ~\mathrm{g} \, ,
\end{eqnarray}
where the coefficient,
\begin{eqnarray}
m_3 &\equiv& \biggl(- \xi^2 \frac{d\theta_3}{d\xi}
\biggr)_{\xi = \xi_1(\theta_3)}
= 2.01824 \, ,
\end{eqnarray}
represents a structural property of $n=3$ polytropes ($\gamma =
4/3$ gases) whose numerical value can be found in Chapter $IV$,
Table 4 or \cite{Chandrasekhar67}. We note as well that
\cite{Chandrasekhar67} identified a characteristic radius,
$\ell_1$, for WDs given by the expression,
\begin{eqnarray}
\ell_1 \mu_e &\equiv& \biggl( \frac{2A}{\pi G} \biggr)^{1/2}
\frac{\mu_e}{B} = 7.71395 \times 10^8 ~\mathrm{cm} \, .
\end{eqnarray}

\subsection{The ``Nauenberg'' Mass-Radius Relationship}

\cite{Nauenberg72} derived an analytic approximation for the
mass-radius relationship exhibited by isolated, spherical WDs
that obey the ZTWD equation of state given in
Eq.~(\ref{eq:EOSchandra}).  Specifically, he offered an
expression of the form,
\begin{eqnarray}
R &=& R_0 \biggl[ \frac{(1 - n^{4/3})^{1/2}}{n^{1/3}} \biggr] \, ,
\label{eq:MRnauenberg}
\end{eqnarray}
where,
\begin{eqnarray}
n &\equiv& \frac{M}{(\mu m_\mu)N_0} \, , \label{eq:nNauenberg} \\
N_0 &\equiv& \frac{(3\pi^2 \zeta)^{1/2}}{\nu^{3/2}} \biggl[
\frac{hc}{2\pi G (\mu m_\mu)^2} \biggr]^{3/2} =
\frac{\mu_e^2 m_p^2}{(\mu m_\mu)^3} \biggl[ \frac{4 \pi \zeta}{m_3^2 \nu^3}
\biggr]^{1/2} M_\mathrm{ch}   \, , \label{eq:N0nauenberg} \\
R_0 &\equiv& (3\pi^2 \zeta)^{1/3} \biggl[ \frac{h}{2\pi m_e c}
\biggr] N_0^{1/3}  = \frac{(\mu_e m_p)}{(\mu m_\mu)} \biggl[ \frac{4 \pi \zeta}{\nu}
\biggr]^{1/2} \ell_1  \, , \label{eq:R0nauenberg}
\end{eqnarray}
$m_\mu$ is the atomic mass unit (see Table
\ref{tab:physicalConstants}), $\mu$ is the mean molecular
weight of the gas, and $\zeta$ and $\nu$ are two adjustable
parameters in Nauenberg's analytic approximation, both of which
are expected to be of order unity. By assuming that the average
particle mass denoted by \cite{Chandrasekhar67} as $(\mu_e
m_p)$ is identical to the average particle mass specified by
\cite{Nauenberg72} as $(\mu m_\mu)$ and, following Nauenberg's
lead, by setting $\nu=1$ and\footnote{Actually,
\cite{Nauenberg72} sets $\zeta=0.323$.},
\begin{eqnarray}
\zeta &=& \frac{m_3^2}{4\pi} = 0.324142 \, ,
\end{eqnarray}
in Eq.~(\ref{eq:N0nauenberg}) we see that,
\begin{eqnarray}
(\mu m_\mu) N_0 &=& M_\mathrm{ch} \, .
\end{eqnarray}
Hence, the denominator in (\ref{eq:nNauenberg}) becomes the
Chandrasekhar mass. Furthermore, expressions
(\ref{eq:R0nauenberg}) and (\ref{eq:MRnauenberg}) become,
respectively,
\begin{eqnarray}
\mu_e R_0 &=& m_3 (\ell_1 \mu_e) = 1.55686 \times 10^{9}~\mathrm{cm} \, ,
\end{eqnarray}
and,
\begin{eqnarray}
R &=& R_0 \biggl\{ \frac{[1 - (M/M_\mathrm{ch})^{4/3}]^{1/2}}{(M/M_\mathrm{ch})^{1/3}}
\biggr\} \, .
\end{eqnarray}
Finally, by adopting the values of $M_\odot$ and $R_\odot$
listed in Table \ref{tab:physicalConstants}, we obtain
essentially\footnote{The numerical coefficients that appear
here in Eqs.~(\ref{eq:MRnauenbergAppendix}) and
(\ref{eq:MchAppendix}) differ slightly from the ones presented
in Eqs.~(27) and (28), respectively, of \cite{Nauenberg72}
presumably because the values of the physical constants ---
such as $M_\odot$ and $R_\odot$ --- that we have adopted in
this paper (see Table \ref{tab:physicalConstants}) are slightly
different from the values adopted by Nauenberg.} the identical
approximate, analytic mass-radius relationship for ZTWDs
presented in Eqs.~(27) and (28) of \cite{Nauenberg72}:
\begin{eqnarray}
\frac{R}{R_\odot} &=& \frac{0.0224}{\mu_e}
\biggl\{ \frac{[1 - (M/M_\mathrm{ch})^{4/3}]^{1/2}}{(M/M_\mathrm{ch})^{1/3}}
\biggr\} \, ,\label{eq:MRnauenbergAppendix}
\end{eqnarray}
where,
\begin{eqnarray}
\frac{M_\mathrm{ch}}{M_\odot} &=& \frac{5.742}{\mu_e^2} \, . \label{eq:MchAppendix}
\end{eqnarray}

\subsection{The ``Eggleton'' Mass-Radius Relationship}

\cite{VR88} introduced the following approximate, analytic
expression for the mass-radius relationship of a ``completely
degenerate $\ldots$ star composed of pure helium'' ({\it i.e.},
$\mu_e=2$), attributing its origin to Eggleton (private
communication):
\begin{eqnarray}
\frac{R}{R_\odot} &=& 0.0114 \biggl[ \biggl( \frac{M}{M_\mathrm{ch}}
\biggr)^{-2/3} -
\biggl( \frac{M}{M_\mathrm{ch}} \biggr)^{2/3} \biggr]^{1/2} \biggl[
1 + 3.5 \biggl( \frac{M}{M_p} \biggr)^{-2/3} + \biggl(
\frac{M}{M_p} \biggr)^{-1} \biggr]^{-2/3}, \label{eq:MReggleton}
\end{eqnarray}
where $M_p$ is a constant whose numerical value is $0.00057
M_\odot$. This ``Eggleton'' mass-radius relationship has been
used widely by researchers when modeling the evolution of
semi-detached binary star systems in which the donor is a ZTWD.
Since the \cite{Nauenberg72} mass-radius relationship
(\ref{eq:MRnauenbergAppendix}) is retrieved from
Eq.~(\ref{eq:MReggleton}) in the limit $M/M_p \gg 1$, it seems
clear that Eggleton's contribution was the insertion of the
term in square brackets involving the ratio $M/M_p$ which, as
\cite{MNS04} phrase it, ``allows for the change to a constant
density configuration at low masses \citep{ZS69}.''  In this
paper we have only constructed binary star systems in which the
internal structure of both stars is governed by the ZTWD
equation of state (\ref{eq:EOSchandra}). Hence it is
appropriate for us to compare the properties of our modeled
systems to behaviors predicted by the ``Nauenberg,'' not the
``Eggleton,'' mass-radius relationship.

\clearpage

\begin{deluxetable}{ccccc}
    \tablewidth{0pt}
    \tablenum{1}
    \tablecolumns{5}
    \tablecaption{Convergence of SCF Method: Binary Model
                  B3 \label{tab:scf_convg9_6}}
    \tablehead{
      \colhead{$N_\varpi$} & \colhead{$N_\theta$} & \colhead{$N_z$} & \colhead{$\delta$} & \colhead{VE} }

    \startdata
    $64$  &  $128$ & $33$ & $1.0 \times 10^{-2}$ & $4.5 \times 10^{-3}$ \\
        &     &     & $1.0 \times 10^{-3}$ & $2.6 \times 10^{-3}$ \\
        &     &     & $1.4 \times 10^{-4}$ & $2.2 \times 10^{-3}$ \\
    $128$ &  $256$ & $65$ & $1.0 \times 10^{-2}$ & $4.0 \times 10^{-3}$ \\
        &     &     & $1.0 \times 10^{-3}$ & $9.1 \times 10^{-4}$ \\
        &     &     & $1.0 \times 10^{-4}$ & $5.7 \times 10^{-4}$ \\
        &     &     & $3.5 \times 10^{-5}$ & $5.4 \times 10^{-4}$
    \enddata
\end{deluxetable}

\begin{deluxetable}{ccccccc}
    \tablewidth{0pt}
    \tablenum{2}
    \tablecolumns{7}
    \tablecaption{Sequence of single, nonrotating ZTWDs. \label{tab:scf_single}}
    \tablehead{
      \colhead{$M$} & \colhead{$R$} & \colhead{$\rho_\mathrm{max}$} & \colhead{$x_\mathrm{max}$}
      & \colhead{$I$} & \colhead{$k$} & \colhead{VE} \\
      ($M_\odot$) & ($10^8$ cm) & ($\mathrm{g}~\mathrm{cm}^{-3}$) &
      & ($10^{50}~\mathrm{g}~\mathrm{cm}^2$) &  &  }

    \startdata
$	0.0844	$&$	19.7673	$&$	3.1623	\times 10^{	4	}	$&$	0.2526	$&$	1.3317	$&$	0.2036	$&$	1.4	\times 10^{	-5	}	$ \\
$	0.1113	$&$	17.9368	$&$	5.6234	\times 10^{	4	}	$&$	0.3060	$&$	1.4422	$&$	0.2031	$&$	6.1	\times 10^{	-5	}	$ \\
$	0.1460	$&$	16.2672	$&$	1.0000	\times 10^{	5	}	$&$	0.3708	$&$	1.5508	$&$	0.2024	$&$	1.4	\times 10^{	-5	}	$ \\
$	0.1903	$&$	14.7421	$&$	1.7783	\times 10^{	5	}	$&$	0.4492	$&$	1.6517	$&$	0.2015	$&$	5.8	\times 10^{	-5	}	$ \\
$	0.2457	$&$	13.3464	$&$	3.1623	\times 10^{	5	}	$&$	0.5442	$&$	1.7360	$&$	0.2001	$&$	1.4	\times 10^{	-5	}	$ \\
$	0.3134	$&$	12.0666	$&$	5.6234	\times 10^{	5	}	$&$	0.6593	$&$	1.7936	$&$	0.1983	$&$	5.8	\times 10^{	-5	}	$ \\
$	0.3938	$&$	10.8906	$&$	1.0000	\times 10^{	6	}	$&$	0.7988	$&$	1.8133	$&$	0.1958	$&$	5.9	\times 10^{	-5	}	$ \\
$	0.4859	$&$	9.8078	$&$	1.7783	\times 10^{	6	}	$&$	0.9678	$&$	1.7852	$&$	0.1926	$&$	1.6	\times 10^{	-5	}	$ \\
$	0.5873	$&$	8.8097	$&$	3.1623	\times 10^{	6	}	$&$	1.1725	$&$	1.7051	$&$	0.1887	$&$	6.1	\times 10^{	-5	}	$ \\
$	0.6942	$&$	7.8896	$&$	5.6234	\times 10^{	6	}	$&$	1.4205	$&$	1.5754	$&$	0.1839	$&$	5.8	\times 10^{	-5	}	$ \\
$	0.8018	$&$	7.0419	$&$	1.0000	\times 10^{	7	}	$&$	1.7209	$&$	1.4058	$&$	0.1783	$&$	7.6	\times 10^{	-5	}	$ \\
$	0.9058	$&$	6.2628	$&$	1.7783	\times 10^{	7	}	$&$	2.0850	$&$	1.2122	$&$	0.1721	$&$	6.6	\times 10^{	-5	}	$ \\
$	1.0022	$&$	5.5487	$&$	3.1623	\times 10^{	7	}	$&$	2.5260	$&$	1.0111	$&$	0.1653	$&$	5.1	\times 10^{	-5	}	$ \\
$	1.0882	$&$	4.8961	$&$	5.6234	\times 10^{	7	}	$&$	3.0603	$&$	0.8176	$&$	0.1581	$&$	6.3	\times 10^{	-5	}	$ \\
$	1.1624	$&$	4.3023	$&$	1.0000	\times 10^{	8	}	$&$	3.7076	$&$	0.6429	$&$	0.1507	$&$	6.3	\times 10^{	-5	}	$ \\
$	1.2246	$&$	3.7643	$&$	1.7783	\times 10^{	8	}	$&$	4.4919	$&$	0.4929	$&$	0.1433	$&$	7.0	\times 10^{	-5	}	$ \\
$	1.2753	$&$	3.2793	$&$	3.1623	\times 10^{	8	}	$&$	5.4421	$&$	0.3698	$&$	0.1360	$&$	7.4	\times 10^{	-5	}	$ \\
$	1.3155	$&$	2.8443	$&$	5.6234	\times 10^{	8	}	$&$	6.5932	$&$	0.2723	$&$	0.1290	$&$	9.0	\times 10^{	-5	}	$ \\
$	1.3469	$&$	2.4560	$&$	1.0000	\times 10^{	9	}	$&$	7.9879	$&$	0.1972	$&$	0.1225	$&$	1.1	\times 10^{	-4	}	$ \\
$	1.3708	$&$	2.1116	$&$	1.7783	\times 10^{	9	}	$&$	9.6775	$&$	0.1410	$&$	0.1164	$&$	1.2	\times 10^{	-4	}	$ \\
$	1.3887	$&$	1.8078	$&$	3.1623	\times 10^{	9	}	$&$	11.7246	$&$	0.0997	$&$	0.1108	$&$	1.3	\times 10^{	-4	}	$ \\
$	1.4020	$&$	1.5414	$&$	5.6234	\times 10^{	9	}	$&$	14.2047	$&$	0.0699	$&$	0.1058	$&$	1.4	\times 10^{	-4	}	$ \\
$	1.4116	$&$	1.3092	$&$	1.0000	\times 10^{	10	}	$&$	17.2094	$&$	0.0486	$&$	0.1013	$&$	1.4	\times 10^{	-4	}	$ \\
    \enddata
\end{deluxetable}

\begin{landscape}
\begin{deluxetable}{cccccccc}
    \tablewidth{0pt}
    \tablenum{3}
    \tablecolumns{8}
    \tablecaption{Selected single, nonrotating ZTWDs. \label{tab:scf_single2}}
    \tablehead{
      \colhead{$M$} & \colhead{$R$} & \colhead{$\rho_\mathrm{max}$} & \colhead{$x_\mathrm{max}$}
       & \colhead{$k$} & \colhead{$W$} & \colhead{$U$} & \colhead{$\Pi$}  \\
      ($M_\odot$) & ($10^8$ cm) & ($\mathrm{g}~\mathrm{cm}^{-3}$) &
       & & ($10^{50}~\mathrm{ergs}$)& ($10^{50}~\mathrm{ergs}$)& ($10^{50}~\mathrm{ergs}$)   }

    \startdata
$	0.5019	$&$	9.6383	$&$	1.9536	\times 10^{	6	}	$&$	0.9979	$&$	0.1923	$&$	-0.6130	$&$	0.3326	$&$	0.2044	$ \\
$	0.6022	$&$	8.6713	$&$	3.4341	\times 10^{	6	}	$&$	1.2043	$&$	0.1889	$&$	-0.9932	$&$	0.5526	$&$	0.3311	$ \\
$	0.7522	$&$	7.4244	$&$	7.6648	\times 10^{	6	}	$&$	1.5739	$&$	0.1813	$&$	-1.8512	$&$	1.0757	$&$	0.6171	$ \\
$	0.9028	$&$	6.2847	$&$	1.7483	\times 10^{	7	}	$&$	2.0718	$&$	0.1726	$&$	-3.2433	$&$	1.9860	$&$	1.0812	$ \\
$	1.0025	$&$	5.5456	$&$	3.1703	\times 10^{	7	}	$&$	2.5265	$&$	0.1655	$&$	-4.6465	$&$	2.9644	$&$	1.5490	$ \\
   \enddata
\end{deluxetable}
\end{landscape}

\clearpage

\begin{deluxetable}{rrrrrrrr}
   \tablenum{4}
   \tablewidth{0pt}
   \tablecolumns{8}
   \tablecaption{DWD Inspiral Sequence `A':
                 $M_\mathrm{tot}=1.5 M_\odot$;
                 $q=1$\label{tab:scf_models_q_1.00}}
   \tablehead{
     \colhead{Model} &
           \colhead{$a$} &
           \colhead{$\Omega$}  &
           \colhead{$M_\mathrm{tot}$}  &
           \colhead{$q$} &
           \colhead{$J_\mathrm{tot}$} &
           \colhead{$E_\mathrm{tot}$}  &
           \colhead{VE} \\
            &
           ($10^9$ cm) &
           ($10^{-2}~\mathrm{s}^{-1})$ &
           ($M_{\sun}$) &
            &
           ($10^{50}~\mathrm{cgs}$) &
           ($10^{50}~\mathrm{erg}$) &
           }

           \startdata
$	A1	$&$	2.0956	$&$	14.8480	$&$	1.5045	$&$	1.0000	$&$	5.3879	$&$	-1.8624	$&$	2.7	 \times 10^{-4}	$ \\
$	A2	$&$	2.0970	$&$	14.8317	$&$	1.5043	$&$	1.0000	$&$	5.3881	$&$	-1.8618	$&$	2.8	 \times 10^{-4}	$ \\
$	A3	$&$	2.1042	$&$	14.7493	$&$	1.5036	$&$	1.0000	$&$	5.3882	$&$	-1.8589	$&$	2.8	 \times 10^{-4}	$ \\
$	A4	$&$	2.1099	$&$	14.6847	$&$	1.5030	$&$	1.0000	$&$	5.3886	$&$	-1.8567	$&$	2.9	 \times 10^{-4}	$ \\
$	A5	$&$	2.1162	$&$	14.6156	$&$	1.5031	$&$	1.0000	$&$	5.3915	$&$	-1.8566	$&$	2.7	 \times 10^{-4}	$ \\
$	A6	$&$	2.1239	$&$	14.5339	$&$	1.5031	$&$	1.0000	$&$	5.3954	$&$	-1.8560	$&$	2.7	 \times 10^{-4}	$ \\
$	A7	$&$	2.1428	$&$	14.3360	$&$	1.5032	$&$	1.0000	$&$	5.4059	$&$	-1.8549	$&$	2.8	 \times 10^{-4}	$ \\
$	A8	$&$	2.1544	$&$	14.2154	$&$	1.5030	$&$	1.0000	$&$	5.4112	$&$	-1.8532	$&$	2.9	 \times 10^{-4}	$ \\
$	A9	$&$	2.1671	$&$	14.0863	$&$	1.5029	$&$	1.0000	$&$	5.4180	$&$	-1.8519	$&$	2.9	 \times 10^{-4}	$ \\
$	A10	$&$	2.1809	$&$	13.9475	$&$	1.5028	$&$	1.0000	$&$	5.4254	$&$	-1.8508	$&$	2.6	 \times 10^{-4}	$ \\
$	A11	$&$	2.1960	$&$	13.7990	$&$	1.5027	$&$	1.0000	$&$	5.4337	$&$	-1.8490	$&$	2.7	 \times 10^{-4}	$ \\
$	A12	$&$	2.2292	$&$	13.4849	$&$	1.5030	$&$	1.0000	$&$	5.4554	$&$	-1.8471	$&$	2.6	 \times 10^{-4}	$ \\
$	A13	$&$	2.2479	$&$	13.3121	$&$	1.5027	$&$	1.0000	$&$	5.4665	$&$	-1.8448	$&$	2.7	 \times 10^{-4}	$ \\
$	A14	$&$	2.2669	$&$	13.1429	$&$	1.5032	$&$	1.0000	$&$	5.4814	$&$	-1.8447	$&$	2.7	 \times 10^{-4}	$ \\
$	A15	$&$	2.2880	$&$	12.9570	$&$	1.5030	$&$	1.0000	$&$	5.4944	$&$	-1.8421	$&$	2.7	 \times 10^{-4}	$ \\
$	A16	$&$	2.3103	$&$	12.7655	$&$	1.5027	$&$	1.0000	$&$	5.5082	$&$	-1.8392	$&$	2.8	 \times 10^{-4}	$ \\
$	A17	$&$	2.3572	$&$	12.3804	$&$	1.5029	$&$	1.0000	$&$	5.5423	$&$	-1.8359	$&$	2.8	 \times 10^{-4}	$ \\
$	A18	$&$	2.3828	$&$	12.1772	$&$	1.5027	$&$	1.0000	$&$	5.5597	$&$	-1.8329	$&$	2.8	 \times 10^{-4}	$ \\
$	A19	$&$	2.4092	$&$	11.9748	$&$	1.5028	$&$	1.0000	$&$	5.5792	$&$	-1.8307	$&$	2.8	 \times 10^{-4}	$ \\
$	A20	$&$	2.4362	$&$	11.7751	$&$	1.5032	$&$	1.0000	$&$	5.6013	$&$	-1.8298	$&$	2.8	 \times 10^{-4}	$ \\
$	A21	$&$	2.4652	$&$	11.5645	$&$	1.5030	$&$	1.0000	$&$	5.6221	$&$	-1.8265	$&$	2.9	 \times 10^{-4}	$ \\
$	A22	$&$	2.5261	$&$	11.1442	$&$	1.5030	$&$	1.0000	$&$	5.6684	$&$	-1.8215	$&$	2.8	 \times 10^{-4}	$ \\
$	A23	$&$	2.5584	$&$	10.9314	$&$	1.5030	$&$	1.0000	$&$	5.6928	$&$	-1.8185	$&$	2.9	 \times 10^{-4}	$ \\
$	A24	$&$	2.5920	$&$	10.7176	$&$	1.5028	$&$	1.0000	$&$	5.7185	$&$	-1.8151	$&$	3.1	 \times 10^{-4}	$ \\
$	A25	$&$	2.6264	$&$	10.5057	$&$	1.5029	$&$	1.0000	$&$	5.7457	$&$	-1.8127	$&$	3.0	 \times 10^{-4}	$ \\
$	A26	$&$	2.6624	$&$	10.2904	$&$	1.5028	$&$	1.0000	$&$	5.7729	$&$	-1.8093	$&$	3.0	 \times 10^{-4}	$ \\
$	A27	$&$	2.6991	$&$	10.0786	$&$	1.5029	$&$	1.0000	$&$	5.8016	$&$	-1.8068	$&$	2.9	 \times 10^{-4}	$ \\
$	A28	$&$	2.7376	$&$	9.8598	$&$	1.5028	$&$	1.0000	$&$	5.8286	$&$	-1.8039	$&$	1.5	 \times 10^{-4}	$ \\
$	A29	$&$	2.8175	$&$	9.4479	$&$	1.5031	$&$	1.0000	$&$	5.8981	$&$	-1.7984	$&$	3.2	 \times 10^{-4}	$ \\
$	A30	$&$	2.8597	$&$	9.2377	$&$	1.5031	$&$	1.0000	$&$	5.9315	$&$	-1.7951	$&$	3.2	 \times 10^{-4}	$ \\
$	A31	$&$	2.8784	$&$	9.1459	$&$	1.5027	$&$	1.0000	$&$	5.9440	$&$	-1.7925	$&$	3.2	 \times 10^{-4}	$ \\
$	A32	$&$	2.9478	$&$	8.8242	$&$	1.5032	$&$	1.0000	$&$	6.0024	$&$	-1.7892	$&$	3.3	 \times 10^{-4}	$ \\
$	A33	$&$	2.9940	$&$	8.6120	$&$	1.5031	$&$	1.0000	$&$	6.0336	$&$	-1.7860	$&$	1.9	 \times 10^{-4}	$ \\
$	A34	$&$	3.0420	$&$	8.4102	$&$	1.5031	$&$	1.0000	$&$	6.0738	$&$	-1.7827	$&$	2.1	 \times 10^{-4}	$ \\
$	A35	$&$	3.0911	$&$	8.2099	$&$	1.5032	$&$	1.0000	$&$	6.1135	$&$	-1.7797	$&$	2.2	 \times 10^{-4}	$ \\
           \enddata
\end{deluxetable}

\begin{landscape}
\begin{deluxetable}{rrrrrrrrrrr}
   \tablenum{5}
   \tablewidth{0pt}
   \tablecolumns{11}
   \tablecaption{Individual Stellar Components along DWD Inspiral
                 Sequence `A'\label{tab:scf_models_ind_q_1.00}}
   \tablehead{
     \colhead{Model} &
           \colhead{$M_1$} &
           \colhead{$R_1$}  &
           \colhead{$\rho^{i=1}_\mathrm{max}$}  &
           \colhead{$f^{i=1}_\mathrm{RL}$} &
           \colhead{$J^{i=1}_\mathrm{spin}$} &
           \colhead{$M_2$} &
           \colhead{$R_2$}  &
           \colhead{$\rho^{i=2}_\mathrm{max}$}  &
           \colhead{$f^{i=2}_\mathrm{RL}$} &
           \colhead{$J^{i=2}_\mathrm{spin}$} \\
            &
           \colhead{($M_{\sun}$)} &
           \colhead{($10^9~\mathrm{cm}$)} &
           \colhead{($10^9~\mathrm{cgs}$)} &
           &
           \colhead{($10^{50}~\mathrm{cgs}$)} &
           \colhead{($M_{\sun}$)} &
           \colhead{($10^9~\mathrm{cm}$)} &
           \colhead{($10^9~\mathrm{cgs}$)} &
           &
           \colhead{($10^{50}~\mathrm{cgs}$)}
           }

           \startdata
$	A1	$&$	0.7522	$&$	0.7841	$&$	6.745	$&$	1.0000	$&$	0.2562	$&$	0.7522	$&$	0.7841	$&$	6.745	$&$	1.0000	$&$	0.2562	$ \\
$	A2	$&$	0.7522	$&$	0.7840	$&$	6.745	$&$	0.9969	$&$	0.2559	$&$	0.7522	$&$	0.7840	$&$	6.745	$&$	0.9969	$&$	0.2559	$ \\
$	A3	$&$	0.7518	$&$	0.7835	$&$	6.745	$&$	0.9838	$&$	0.2540	$&$	0.7518	$&$	0.7835	$&$	6.745	$&$	0.9838	$&$	0.2540	$ \\
$	A4	$&$	0.7515	$&$	0.7831	$&$	6.745	$&$	0.9736	$&$	0.2526	$&$	0.7515	$&$	0.7831	$&$	6.745	$&$	0.9736	$&$	0.2526	$ \\
$	A5	$&$	0.7516	$&$	0.7825	$&$	6.758	$&$	0.9614	$&$	0.2509	$&$	0.7516	$&$	0.7825	$&$	6.758	$&$	0.9615	$&$	0.2509	$ \\
$	A6	$&$	0.7516	$&$	0.7817	$&$	6.771	$&$	0.9471	$&$	0.2489	$&$	0.7516	$&$	0.7817	$&$	6.771	$&$	0.9471	$&$	0.2489	$ \\
$	A7	$&$	0.7516	$&$	0.7798	$&$	6.805	$&$	0.9134	$&$	0.2441	$&$	0.7516	$&$	0.7798	$&$	6.805	$&$	0.9134	$&$	0.2441	$ \\
$	A8	$&$	0.7515	$&$	0.7790	$&$	6.820	$&$	0.8942	$&$	0.2414	$&$	0.7515	$&$	0.7790	$&$	6.820	$&$	0.8942	$&$	0.2414	$ \\
$	A9	$&$	0.7514	$&$	0.7780	$&$	6.837	$&$	0.8739	$&$	0.2384	$&$	0.7514	$&$	0.7780	$&$	6.837	$&$	0.8739	$&$	0.2384	$ \\
$	A10	$&$	0.7514	$&$	0.7769	$&$	6.855	$&$	0.8523	$&$	0.2353	$&$	0.7514	$&$	0.7769	$&$	6.855	$&$	0.8523	$&$	0.2353	$ \\
$	A11	$&$	0.7513	$&$	0.7759	$&$	6.876	$&$	0.8303	$&$	0.2320	$&$	0.7513	$&$	0.7758	$&$	6.876	$&$	0.8303	$&$	0.2320	$ \\
$	A12	$&$	0.7515	$&$	0.7735	$&$	6.924	$&$	0.7838	$&$	0.2250	$&$	0.7515	$&$	0.7735	$&$	6.924	$&$	0.7838	$&$	0.2250	$ \\
$	A13	$&$	0.7514	$&$	0.7724	$&$	6.944	$&$	0.7602	$&$	0.2213	$&$	0.7514	$&$	0.7724	$&$	6.944	$&$	0.7602	$&$	0.2213	$ \\
$	A14	$&$	0.7516	$&$	0.7711	$&$	6.977	$&$	0.7367	$&$	0.2176	$&$	0.7516	$&$	0.7711	$&$	6.977	$&$	0.7367	$&$	0.2176	$ \\
$	A15	$&$	0.7515	$&$	0.7701	$&$	6.992	$&$	0.7131	$&$	0.2138	$&$	0.7515	$&$	0.7701	$&$	6.992	$&$	0.7131	$&$	0.2138	$ \\
$	A16	$&$	0.7513	$&$	0.7692	$&$	7.013	$&$	0.6903	$&$	0.2099	$&$	0.7513	$&$	0.7692	$&$	7.013	$&$	0.6903	$&$	0.2099	$ \\
$	A17	$&$	0.7515	$&$	0.7668	$&$	7.062	$&$	0.6440	$&$	0.2020	$&$	0.7515	$&$	0.7668	$&$	7.062	$&$	0.6440	$&$	0.2020	$ \\
$	A18	$&$	0.7514	$&$	0.7658	$&$	7.083	$&$	0.6215	$&$	0.1980	$&$	0.7514	$&$	0.7658	$&$	7.083	$&$	0.6215	$&$	0.1980	$ \\
$	A19	$&$	0.7514	$&$	0.7650	$&$	7.103	$&$	0.5998	$&$	0.1940	$&$	0.7514	$&$	0.7650	$&$	7.103	$&$	0.5998	$&$	0.1940	$ \\
$	A20	$&$	0.7516	$&$	0.7637	$&$	7.137	$&$	0.5779	$&$	0.1901	$&$	0.7516	$&$	0.7637	$&$	7.137	$&$	0.5779	$&$	0.1901	$ \\
$	A21	$&$	0.7515	$&$	0.7629	$&$	7.153	$&$	0.5568	$&$	0.1861	$&$	0.7515	$&$	0.7629	$&$	7.153	$&$	0.5567	$&$	0.1861	$ \\
$	A22	$&$	0.7515	$&$	0.7611	$&$	7.198	$&$	0.5160	$&$	0.1781	$&$	0.7515	$&$	0.7611	$&$	7.198	$&$	0.5160	$&$	0.1781	$ \\
$	A23	$&$	0.7515	$&$	0.7603	$&$	7.214	$&$	0.4964	$&$	0.1742	$&$	0.7515	$&$	0.7603	$&$	7.214	$&$	0.4964	$&$	0.1742	$ \\
$	A24	$&$	0.7514	$&$	0.7596	$&$	7.234	$&$	0.4774	$&$	0.1703	$&$	0.7514	$&$	0.7596	$&$	7.234	$&$	0.4774	$&$	0.1703	$ \\
$	A25	$&$	0.7515	$&$	0.7585	$&$	7.251	$&$	0.4583	$&$	0.1664	$&$	0.7515	$&$	0.7585	$&$	7.251	$&$	0.4583	$&$	0.1664	$ \\
$	A26	$&$	0.7514	$&$	0.7579	$&$	7.271	$&$	0.4404	$&$	0.1625	$&$	0.7514	$&$	0.7579	$&$	7.271	$&$	0.4404	$&$	0.1625	$ \\
$	A27	$&$	0.7514	$&$	0.7571	$&$	7.290	$&$	0.4231	$&$	0.1587	$&$	0.7514	$&$	0.7571	$&$	7.290	$&$	0.4231	$&$	0.1587	$ \\
$	A28	$&$	0.7514	$&$	0.7562	$&$	7.309	$&$	0.4057	$&$	0.1548	$&$	0.7514	$&$	0.7562	$&$	7.309	$&$	0.4057	$&$	0.1548	$ \\
$	A29	$&$	0.7516	$&$	0.7548	$&$	7.349	$&$	0.3733	$&$	0.1476	$&$	0.7516	$&$	0.7548	$&$	7.349	$&$	0.3733	$&$	0.1476	$ \\
$	A30	$&$	0.7515	$&$	0.7541	$&$	7.363	$&$	0.3575	$&$	0.1440	$&$	0.7515	$&$	0.7541	$&$	7.363	$&$	0.3576	$&$	0.1440	$ \\
$	A31	$&$	0.7513	$&$	0.7540	$&$	7.363	$&$	0.3478	$&$	0.1424	$&$	0.7513	$&$	0.7541	$&$	7.363	$&$	0.3518	$&$	0.1424	$ \\
$	A32	$&$	0.7516	$&$	0.7529	$&$	7.397	$&$	0.3282	$&$	0.1369	$&$	0.7516	$&$	0.7529	$&$	7.397	$&$	0.3282	$&$	0.1369	$ \\
$	A33	$&$	0.7515	$&$	0.7522	$&$	7.412	$&$	0.3142	$&$	0.1333	$&$	0.7515	$&$	0.7522	$&$	7.412	$&$	0.3142	$&$	0.1333	$ \\
$	A34	$&$	0.7516	$&$	0.7517	$&$	7.425	$&$	0.3005	$&$	0.1299	$&$	0.7516	$&$	0.7517	$&$	7.425	$&$	0.3005	$&$	0.1299	$ \\
$	A35	$&$	0.7516	$&$	0.7511	$&$	7.438	$&$	0.2874	$&$	0.1265	$&$	0.7516	$&$	0.7510	$&$	7.440	$&$	0.2873	$&$	0.1265	$ \\
           \enddata
\end{deluxetable}
\end{landscape}

\begin{deluxetable}{rrrrrrrr}
   \tablenum{6}
   \tablewidth{0pt}
   \tablecolumns{8}
   \tablecaption{DWD Inspiral Sequence `B':
                 $M_\mathrm{tot}=1.5 M_\odot$;
                 $q=2/3$\label{tab:scf_models_q_0.66}}
   \tablehead{
     \colhead{Model} &
           \colhead{$a$} &
           \colhead{$\Omega$}  &
           \colhead{$M_\mathrm{tot}$}  &
           \colhead{$q$} &
           \colhead{$J_\mathrm{tot}$} &
           \colhead{$E_\mathrm{tot}$}  &
           \colhead{VE} \\
            &
           ($10^9$ cm) &
           ($10^{-2}~\mathrm{s}^{-1})$ &
           ($M_{\sun}$) &
            &
           ($10^{50}~\mathrm{cgs}$) &
           ($10^{50}~\mathrm{erg}$) &
           }

           \startdata
$	B1	$&$	2.6679	$&$	10.2944	$&$	1.5042	$&$	0.6671	$&$	5.5888	$&$	-1.9460	$&$	6.0	 \times 10^{-4}	$ \\
$	B2	$&$	2.6743	$&$	10.2576	$&$	1.5043	$&$	0.6667	$&$	5.5931	$&$	-1.9460	$&$	6.1	 \times 10^{-4}	$ \\
$	B3	$&$	2.6819	$&$	10.2106	$&$	1.5037	$&$	0.6667	$&$	5.5951	$&$	-1.9436	$&$	6.0	 \times 10^{-4}	$ \\
$	B4	$&$	2.6923	$&$	10.1491	$&$	1.5033	$&$	0.6664	$&$	5.6001	$&$	-1.9419	$&$	6.1	 \times 10^{-4}	$ \\
$	B5	$&$	2.7039	$&$	10.0831	$&$	1.5036	$&$	0.6663	$&$	5.6090	$&$	-1.9420	$&$	6.0	 \times 10^{-4}	$ \\
$	B6	$&$	2.7177	$&$	10.0040	$&$	1.5034	$&$	0.6665	$&$	5.6174	$&$	-1.9401	$&$	6.1	 \times 10^{-4}	$ \\
$	B7	$&$	2.7487	$&$	9.8330	$&$	1.5035	$&$	0.6665	$&$	5.6403	$&$	-1.9383	$&$	6.2	 \times 10^{-4}	$ \\
$	B8	$&$	2.7687	$&$	9.7249	$&$	1.5035	$&$	0.6664	$&$	5.6542	$&$	-1.9372	$&$	6.2	 \times 10^{-4}	$ \\
$	B9	$&$	2.7902	$&$	9.6104	$&$	1.5033	$&$	0.6665	$&$	5.6684	$&$	-1.9348	$&$	6.3	 \times 10^{-4}	$ \\
$	B10	$&$	2.8138	$&$	9.4879	$&$	1.5033	$&$	0.6664	$&$	5.6851	$&$	-1.9333	$&$	6.3	 \times 10^{-4}	$ \\
$	B11	$&$	2.8388	$&$	9.3609	$&$	1.5033	$&$	0.6663	$&$	5.7034	$&$	-1.9318	$&$	6.4	 \times 10^{-4}	$ \\
$	B12	$&$	2.8654	$&$	9.2290	$&$	1.5032	$&$	0.6665	$&$	5.7224	$&$	-1.9295	$&$	6.4	 \times 10^{-4}	$ \\
$	B13	$&$	2.9196	$&$	8.9712	$&$	1.5034	$&$	0.6665	$&$	5.7638	$&$	-1.9267	$&$	6.5	 \times 10^{-4}	$ \\
$	B14	$&$	2.9514	$&$	8.8246	$&$	1.5033	$&$	0.6665	$&$	5.7864	$&$	-1.9241	$&$	6.8	 \times 10^{-4}	$ \\
$	B15	$&$	2.9850	$&$	8.6746	$&$	1.5031	$&$	0.6665	$&$	5.8108	$&$	-1.9215	$&$	6.8	 \times 10^{-4}	$ \\
$	B16	$&$	3.0199	$&$	8.5240	$&$	1.5034	$&$	0.6664	$&$	5.8380	$&$	-1.9201	$&$	6.9	 \times 10^{-4}	$ \\
$	B17	$&$	3.0567	$&$	8.3680	$&$	1.5031	$&$	0.6665	$&$	5.8638	$&$	-1.9170	$&$	7.0	 \times 10^{-4}	$ \\
$	B18	$&$	3.0949	$&$	8.2123	$&$	1.5032	$&$	0.6665	$&$	5.8928	$&$	-1.9147	$&$	7.0	 \times 10^{-4}	$ \\
$	B19	$&$	3.1350	$&$	8.0548	$&$	1.5034	$&$	0.6663	$&$	5.9238	$&$	-1.9132	$&$	7.2	 \times 10^{-4}	$ \\
$	B20	$&$	3.1766	$&$	7.8955	$&$	1.5031	$&$	0.6665	$&$	5.9543	$&$	-1.9097	$&$	7.4	 \times 10^{-4}	$ \\
$	B21	$&$	3.2199	$&$	7.7355	$&$	1.5032	$&$	0.6666	$&$	5.9870	$&$	-1.9074	$&$	7.5	 \times 10^{-4}	$ \\
$	B22	$&$	3.2653	$&$	7.5743	$&$	1.5034	$&$	0.6664	$&$	6.0218	$&$	-1.9056	$&$	7.8	 \times 10^{-4}	$ \\
$	B23	$&$	3.3554	$&$	7.2699	$&$	1.5033	$&$	0.6667	$&$	6.0905	$&$	-1.9002	$&$	8.1	 \times 10^{-4}	$ \\
$	B24	$&$	3.4062	$&$	7.1070	$&$	1.5034	$&$	0.6666	$&$	6.1285	$&$	-1.8978	$&$	8.1	 \times 10^{-4}	$ \\
$	B25	$&$	3.4590	$&$	6.9440	$&$	1.5034	$&$	0.6665	$&$	6.1681	$&$	-1.8951	$&$	8.5	 \times 10^{-4}	$ \\
$	B26	$&$	3.5133	$&$	6.7821	$&$	1.5032	$&$	0.6666	$&$	6.2078	$&$	-1.8917	$&$	8.6	 \times 10^{-4}	$ \\
$	B27	$&$	3.5700	$&$	6.6211	$&$	1.5033	$&$	0.6665	$&$	6.2506	$&$	-1.8893	$&$	8.8	 \times 10^{-4}	$ \\
$	B28	$&$	3.6285	$&$	6.4611	$&$	1.5034	$&$	0.6665	$&$	6.2948	$&$	-1.8867	$&$	9.2	 \times 10^{-4}	$ \\
$	B29	$&$	3.6892	$&$	6.3010	$&$	1.5032	$&$	0.6666	$&$	6.3387	$&$	-1.8832	$&$	9.3	 \times 10^{-4}	$ \\
$	B30	$&$	3.7520	$&$	6.1432	$&$	1.5033	$&$	0.6666	$&$	6.3860	$&$	-1.8809	$&$	9.6	 \times 10^{-4}	$ \\
$	B31	$&$	3.8171	$&$	5.9860	$&$	1.5034	$&$	0.6665	$&$	6.4340	$&$	-1.8783	$&$	9.9	 \times 10^{-4}	$ \\
$	B32	$&$	3.8846	$&$	5.8298	$&$	1.5032	$&$	0.6667	$&$	6.4828	$&$	-1.8747	$&$	1.0	 \times 10^{-3}	$ \\
$	B33	$&$	3.9544	$&$	5.6759	$&$	1.5034	$&$	0.6666	$&$	6.5344	$&$	-1.8724	$&$	1.0	 \times 10^{-3}	$ \\
$	B34	$&$	4.0270	$&$	5.5225	$&$	1.5034	$&$	0.6665	$&$	6.5874	$&$	-1.8698	$&$	1.1	 \times 10^{-3}	$ \\
$	B35	$&$	4.1030	$&$	5.3681	$&$	1.5027	$&$	0.6666	$&$	6.6380	$&$	-1.8645	$&$	1.1	 \times 10^{-3}	$ \\
$	B36	$&$	4.1801	$&$	5.2199	$&$	1.5031	$&$	0.6666	$&$	6.6951	$&$	-1.8628	$&$	1.1	 \times 10^{-3}	$ \\
$	B37	$&$	4.2601	$&$	5.0741	$&$	1.5034	$&$	0.6666	$&$	6.7548	$&$	-1.8611	$&$	1.2	 \times 10^{-3}	$ \\
$	B38	$&$	4.2667	$&$	5.0582	$&$	1.5030	$&$	0.6668	$&$	6.7529	$&$	-1.8596	$&$	1.1	 \times 10^{-3}	$ \\
$	B39	$&$	4.3502	$&$	4.9129	$&$	1.5031	$&$	0.6667	$&$	6.8124	$&$	-1.8570	$&$	1.2	 \times 10^{-3}	$ \\
$	B40	$&$	4.4375	$&$	4.7683	$&$	1.5032	$&$	0.6664	$&$	6.8739	$&$	-1.8548	$&$	1.2	 \times 10^{-3}	$ \\
$	B41	$&$	4.5270	$&$	4.6269	$&$	1.5030	$&$	0.6666	$&$	6.9357	$&$	-1.8513	$&$	1.3	 \times 10^{-3}	$ \\
$	B42	$&$	4.6192	$&$	4.4889	$&$	1.5032	$&$	0.6668	$&$	7.0011	$&$	-1.8487	$&$	1.3	 \times 10^{-3}	$ \\
$	B43	$&$	4.7160	$&$	4.3508	$&$	1.5034	$&$	0.6666	$&$	7.0680	$&$	-1.8467	$&$	1.4	 \times 10^{-3}	$ \\
$	B44	$&$	4.8164	$&$	4.2143	$&$	1.5031	$&$	0.6667	$&$	7.1344	$&$	-1.8429	$&$	1.4	 \times 10^{-3}	$ \\
$	B45	$&$	4.9203	$&$	4.0808	$&$	1.5029	$&$	0.6668	$&$	7.2039	$&$	-1.8395	$&$	1.5	 \times 10^{-3}	$ \\
$	B46	$&$	5.0281	$&$	3.9494	$&$	1.5029	$&$	0.6668	$&$	7.2752	$&$	-1.8366	$&$	1.5	 \times 10^{-3}	$ \\
$	B47	$&$	5.1405	$&$	3.8227	$&$	1.5033	$&$	0.6665	$&$	7.3562	$&$	-1.8352	$&$	1.7	 \times 10^{-3}	$ \\
$	B48	$&$	5.1467	$&$	3.8136	$&$	1.5032	$&$	0.6669	$&$	7.3567	$&$	-1.8347	$&$	1.6	 \times 10^{-3}	$ \\
$	B49	$&$	5.2630	$&$	3.6870	$&$	1.5033	$&$	0.6670	$&$	7.4330	$&$	-1.8320	$&$	1.7	 \times 10^{-3}	$ \\
           \enddata
\end{deluxetable}

\begin{landscape}
\begin{deluxetable}{rrrrrrrrrrr}
   \tablenum{7}
   \tablewidth{0pt}
   \tablecolumns{11}
   \tablecaption{Individual Stellar Components along DWD Inspiral
                 Sequence `B'\label{tab:scf_models_ind_q_0.66}}
   \tablehead{
     \colhead{Model} &
           \colhead{$M_1$} &
           \colhead{$R_1$}  &
           \colhead{$\rho^{i=1}_\mathrm{max}$}  &
           \colhead{$f^{i=1}_\mathrm{RL}$} &
           \colhead{$J^{i=1}_\mathrm{spin}$} &
           \colhead{$M_2$} &
           \colhead{$R_2$}  &
           \colhead{$\rho^{i=2}_\mathrm{max}$}  &
           \colhead{$f^{i=2}_\mathrm{RL}$} &
           \colhead{$J^{i=2}_\mathrm{spin}$} \\
            &
           \colhead{($M_{\sun}$)} &
           \colhead{($10^9~\mathrm{cm}$)} &
           \colhead{($10^9~\mathrm{cgs}$)} &
           &
           \colhead{($10^{50}~\mathrm{cgs}$)} &
           \colhead{($M_{\sun}$)} &
           \colhead{($10^9~\mathrm{cm}$)} &
           \colhead{($10^9~\mathrm{cgs}$)} &
           &
           \colhead{($10^{50}~\mathrm{cgs}$)}
           }

           \startdata
$	B1	$&$	0.9023	$&$	0.6345	$&$	17.049	$&$	0.1913	$&$	0.1289	$&$	0.6019	$&$	0.9117	$&$	3.076	$&$	1.0001	$&$	0.2004	$ \\
$	B2	$&$	0.9026	$&$	0.6344	$&$	17.049	$&$	0.1898	$&$	0.1283	$&$	0.6017	$&$	0.9113	$&$	3.076	$&$	0.9908	$&$	0.1995	$ \\
$	B3	$&$	0.9022	$&$	0.6345	$&$	17.049	$&$	0.1881	$&$	0.1278	$&$	0.6015	$&$	0.9108	$&$	3.076	$&$	0.9805	$&$	0.1982	$ \\
$	B4	$&$	0.9022	$&$	0.6343	$&$	17.049	$&$	0.1856	$&$	0.1270	$&$	0.6012	$&$	0.9101	$&$	3.076	$&$	0.9667	$&$	0.1966	$ \\
$	B5	$&$	0.9023	$&$	0.6343	$&$	17.047	$&$	0.1832	$&$	0.1261	$&$	0.6013	$&$	0.9090	$&$	3.083	$&$	0.9504	$&$	0.1948	$ \\
$	B6	$&$	0.9021	$&$	0.6343	$&$	17.059	$&$	0.1803	$&$	0.1250	$&$	0.6013	$&$	0.9078	$&$	3.089	$&$	0.9314	$&$	0.1927	$ \\
$	B7	$&$	0.9022	$&$	0.6340	$&$	17.082	$&$	0.1737	$&$	0.1228	$&$	0.6013	$&$	0.9056	$&$	3.108	$&$	0.8916	$&$	0.1882	$ \\
$	B8	$&$	0.9022	$&$	0.6336	$&$	17.085	$&$	0.1694	$&$	0.1213	$&$	0.6012	$&$	0.9043	$&$	3.115	$&$	0.8677	$&$	0.1854	$ \\
$	B9	$&$	0.9020	$&$	0.6336	$&$	17.100	$&$	0.1654	$&$	0.1198	$&$	0.6012	$&$	0.9029	$&$	3.123	$&$	0.8430	$&$	0.1825	$ \\
$	B10	$&$	0.9021	$&$	0.6334	$&$	17.112	$&$	0.1610	$&$	0.1182	$&$	0.6011	$&$	0.9017	$&$	3.134	$&$	0.8181	$&$	0.1795	$ \\
$	B11	$&$	0.9022	$&$	0.6334	$&$	17.121	$&$	0.1565	$&$	0.1165	$&$	0.6011	$&$	0.9003	$&$	3.144	$&$	0.7917	$&$	0.1764	$ \\
$	B12	$&$	0.9020	$&$	0.6332	$&$	17.139	$&$	0.1520	$&$	0.1148	$&$	0.6012	$&$	0.8987	$&$	3.153	$&$	0.7650	$&$	0.1732	$ \\
$	B13	$&$	0.9022	$&$	0.6329	$&$	17.162	$&$	0.1432	$&$	0.1114	$&$	0.6013	$&$	0.8963	$&$	3.175	$&$	0.7172	$&$	0.1670	$ \\
$	B14	$&$	0.9021	$&$	0.6326	$&$	17.177	$&$	0.1383	$&$	0.1095	$&$	0.6012	$&$	0.8950	$&$	3.184	$&$	0.6907	$&$	0.1636	$ \\
$	B15	$&$	0.9020	$&$	0.6324	$&$	17.195	$&$	0.1335	$&$	0.1076	$&$	0.6011	$&$	0.8935	$&$	3.194	$&$	0.6644	$&$	0.1602	$ \\
$	B16	$&$	0.9022	$&$	0.6323	$&$	17.203	$&$	0.1288	$&$	0.1056	$&$	0.6012	$&$	0.8925	$&$	3.203	$&$	0.6403	$&$	0.1568	$ \\
$	B17	$&$	0.9020	$&$	0.6323	$&$	17.222	$&$	0.1241	$&$	0.1036	$&$	0.6011	$&$	0.8914	$&$	3.213	$&$	0.6158	$&$	0.1534	$ \\
$	B18	$&$	0.9020	$&$	0.6318	$&$	17.241	$&$	0.1192	$&$	0.1016	$&$	0.6012	$&$	0.8902	$&$	3.223	$&$	0.5916	$&$	0.1500	$ \\
$	B19	$&$	0.9022	$&$	0.6316	$&$	17.246	$&$	0.1145	$&$	0.0995	$&$	0.6012	$&$	0.8887	$&$	3.233	$&$	0.5675	$&$	0.1465	$ \\
$	B20	$&$	0.9019	$&$	0.6315	$&$	17.268	$&$	0.1099	$&$	0.0975	$&$	0.6012	$&$	0.8879	$&$	3.242	$&$	0.5448	$&$	0.1431	$ \\
$	B21	$&$	0.9020	$&$	0.6311	$&$	17.287	$&$	0.1053	$&$	0.0955	$&$	0.6012	$&$	0.8866	$&$	3.251	$&$	0.5219	$&$	0.1397	$ \\
$	B22	$&$	0.9022	$&$	0.6311	$&$	17.294	$&$	0.1009	$&$	0.0933	$&$	0.6012	$&$	0.8856	$&$	3.261	$&$	0.5003	$&$	0.1364	$ \\
$	B23	$&$	0.9020	$&$	0.6307	$&$	17.340	$&$	0.0928	$&$	0.0895	$&$	0.6013	$&$	0.8837	$&$	3.280	$&$	0.4605	$&$	0.1301	$ \\
$	B24	$&$	0.9021	$&$	0.6304	$&$	17.353	$&$	0.0885	$&$	0.0874	$&$	0.6013	$&$	0.8829	$&$	3.285	$&$	0.4408	$&$	0.1268	$ \\
$	B25	$&$	0.9021	$&$	0.6302	$&$	17.362	$&$	0.0844	$&$	0.0853	$&$	0.6012	$&$	0.8823	$&$	3.294	$&$	0.4219	$&$	0.1235	$ \\
$	B26	$&$	0.9019	$&$	0.6303	$&$	17.384	$&$	0.0806	$&$	0.0833	$&$	0.6013	$&$	0.8814	$&$	3.301	$&$	0.4029	$&$	0.1203	$ \\
$	B27	$&$	0.9021	$&$	0.6301	$&$	17.398	$&$	0.0767	$&$	0.0812	$&$	0.6012	$&$	0.8804	$&$	3.309	$&$	0.3846	$&$	0.1171	$ \\
$	B28	$&$	0.9021	$&$	0.6297	$&$	17.409	$&$	0.0729	$&$	0.0792	$&$	0.6013	$&$	0.8791	$&$	3.316	$&$	0.3661	$&$	0.1140	$ \\
$	B29	$&$	0.9019	$&$	0.6298	$&$	17.432	$&$	0.0694	$&$	0.0772	$&$	0.6013	$&$	0.8790	$&$	3.323	$&$	0.3500	$&$	0.1109	$ \\
$	B30	$&$	0.9021	$&$	0.6292	$&$	17.445	$&$	0.0657	$&$	0.0752	$&$	0.6013	$&$	0.8780	$&$	3.330	$&$	0.3334	$&$	0.1078	$ \\
$	B31	$&$	0.9021	$&$	0.6293	$&$	17.457	$&$	0.0624	$&$	0.0732	$&$	0.6013	$&$	0.8772	$&$	3.336	$&$	0.3174	$&$	0.1048	$ \\
$	B32	$&$	0.9019	$&$	0.6288	$&$	17.480	$&$	0.0591	$&$	0.0712	$&$	0.6013	$&$	0.8768	$&$	3.343	$&$	0.3024	$&$	0.1019	$ \\
$	B33	$&$	0.9020	$&$	0.6289	$&$	17.495	$&$	0.0560	$&$	0.0693	$&$	0.6013	$&$	0.8762	$&$	3.349	$&$	0.2879	$&$	0.0990	$ \\
$	B34	$&$	0.9021	$&$	0.6291	$&$	17.504	$&$	0.0531	$&$	0.0674	$&$	0.6013	$&$	0.8753	$&$	3.355	$&$	0.2734	$&$	0.0961	$ \\
$	B35	$&$	0.9017	$&$	0.6287	$&$	17.504	$&$	0.0501	$&$	0.0655	$&$	0.6011	$&$	0.8749	$&$	3.355	$&$	0.2597	$&$	0.0932	$ \\
$	B36	$&$	0.9019	$&$	0.6285	$&$	17.531	$&$	0.0473	$&$	0.0636	$&$	0.6012	$&$	0.8746	$&$	3.365	$&$	0.2469	$&$	0.0905	$ \\
$	B37	$&$	0.9021	$&$	0.6283	$&$	17.542	$&$	0.0447	$&$	0.0617	$&$	0.6013	$&$	0.8739	$&$	3.371	$&$	0.2344	$&$	0.0878	$ \\
$	B38	$&$	0.9017	$&$	0.6284	$&$	17.542	$&$	0.0445	$&$	0.0616	$&$	0.6013	$&$	0.8739	$&$	3.371	$&$	0.2334	$&$	0.0875	$ \\
$	B39	$&$	0.9018	$&$	0.6280	$&$	17.572	$&$	0.0419	$&$	0.0598	$&$	0.6013	$&$	0.8732	$&$	3.378	$&$	0.2212	$&$	0.0848	$ \\
$	B40	$&$	0.9021	$&$	0.6276	$&$	17.572	$&$	0.0394	$&$	0.0579	$&$	0.6011	$&$	0.8728	$&$	3.378	$&$	0.2096	$&$	0.0822	$ \\
$	B41	$&$	0.9018	$&$	0.6273	$&$	17.601	$&$	0.0371	$&$	0.0562	$&$	0.6012	$&$	0.8723	$&$	3.387	$&$	0.1984	$&$	0.0796	$ \\
$	B42	$&$	0.9018	$&$	0.6274	$&$	17.631	$&$	0.0350	$&$	0.0545	$&$	0.6014	$&$	0.8712	$&$	3.392	$&$	0.1875	$&$	0.0771	$ \\
$	B43	$&$	0.9021	$&$	0.6267	$&$	17.645	$&$	0.0328	$&$	0.0527	$&$	0.6013	$&$	0.8710	$&$	3.398	$&$	0.1774	$&$	0.0746	$ \\
$	B44	$&$	0.9019	$&$	0.6272	$&$	17.645	$&$	0.0309	$&$	0.0510	$&$	0.6012	$&$	0.8709	$&$	3.398	$&$	0.1677	$&$	0.0722	$ \\
$	B45	$&$	0.9017	$&$	0.6272	$&$	17.682	$&$	0.0290	$&$	0.0494	$&$	0.6012	$&$	0.8704	$&$	3.407	$&$	0.1583	$&$	0.0698	$ \\
$	B46	$&$	0.9017	$&$	0.6265	$&$	17.682	$&$	0.0271	$&$	0.0478	$&$	0.6012	$&$	0.8699	$&$	3.407	$&$	0.1491	$&$	0.0674	$ \\
$	B47	$&$	0.9020	$&$	0.6266	$&$	17.697	$&$	0.0254	$&$	0.0462	$&$	0.6012	$&$	0.8696	$&$	3.416	$&$	0.1405	$&$	0.0652	$ \\
$	B48	$&$	0.9018	$&$	0.6266	$&$	17.726	$&$	0.0254	$&$	0.0461	$&$	0.6014	$&$	0.8695	$&$	3.419	$&$	0.1400	$&$	0.0650	$ \\
$	B49	$&$	0.9018	$&$	0.6260	$&$	17.758	$&$	0.0237	$&$	0.0445	$&$	0.6015	$&$	0.8691	$&$	3.421	$&$	0.1318	$&$	0.0628	$ \\
           \enddata
\end{deluxetable}
\end{landscape}

\begin{deluxetable}{rrrrrrrr}
   \tablenum{8}
   \tablewidth{0pt}
   \tablecolumns{8}
   \tablecaption{DWD Inspiral Sequence `C':
                 $M_\mathrm{tot}=1.5 M_\odot$;
                 $q=1/2$\label{tab:scf_models_q_0.50}}
   \tablehead{
     \colhead{Model} &
           \colhead{$a$} &
           \colhead{$\Omega$}  &
           \colhead{$M_\mathrm{tot}$}  &
           \colhead{$q$} &
           \colhead{$J_\mathrm{tot}$} &
           \colhead{$E_\mathrm{tot}$}  &
           \colhead{VE} \\
            &
           ($10^9$ cm) &
           ($10^{-2}~\mathrm{s}^{-1})$ &
           ($M_{\sun}$) &
            &
           ($10^{50}~\mathrm{cgs}$) &
           ($10^{50}~\mathrm{erg}$) &
           }

           \startdata
$	C1	$&$	3.1807	$&$	7.9054	$&$	1.5043	$&$	0.5001	$&$	5.5568	$&$	-2.1635	$&$	1.2	 \times 10^{-3}	$ \\
$	C2	$&$	3.1814	$&$	7.9025	$&$	1.5044	$&$	0.5001	$&$	5.5574	$&$	-2.1638	$&$	1.2	 \times 10^{-3}	$ \\
$	C3	$&$	3.1865	$&$	7.8822	$&$	1.5039	$&$	0.5002	$&$	5.5582	$&$	-2.1615	$&$	1.2	 \times 10^{-3}	$ \\
$	C4	$&$	3.2070	$&$	7.8041	$&$	1.5033	$&$	0.4997	$&$	5.5671	$&$	-2.1594	$&$	1.2	 \times 10^{-3}	$ \\
$	C5	$&$	3.2208	$&$	7.7539	$&$	1.5038	$&$	0.4997	$&$	5.5780	$&$	-2.1605	$&$	1.2	 \times 10^{-3}	$ \\
$	C6	$&$	3.2376	$&$	7.6922	$&$	1.5036	$&$	0.4998	$&$	5.5883	$&$	-2.1584	$&$	1.2	 \times 10^{-3}	$ \\
$	C7	$&$	3.2573	$&$	7.6213	$&$	1.5034	$&$	0.4999	$&$	5.6009	$&$	-2.1566	$&$	1.2	 \times 10^{-3}	$ \\
$	C8	$&$	3.2800	$&$	7.5404	$&$	1.5035	$&$	0.4998	$&$	5.6150	$&$	-2.1561	$&$	1.2	 \times 10^{-3}	$ \\
$	C9	$&$	3.3051	$&$	7.4547	$&$	1.5038	$&$	0.4996	$&$	5.6329	$&$	-2.1566	$&$	1.2	 \times 10^{-3}	$ \\
$	C10	$&$	3.3321	$&$	7.3618	$&$	1.5033	$&$	0.4999	$&$	5.6489	$&$	-2.1529	$&$	1.2	 \times 10^{-3}	$ \\
$	C11	$&$	3.3617	$&$	7.2631	$&$	1.5032	$&$	0.4999	$&$	5.6680	$&$	-2.1511	$&$	1.2	 \times 10^{-3}	$ \\
$	C12	$&$	3.4235	$&$	7.0659	$&$	1.5036	$&$	0.4999	$&$	5.7108	$&$	-2.1497	$&$	1.3	 \times 10^{-3}	$ \\
$	C13	$&$	3.4601	$&$	6.9530	$&$	1.5036	$&$	0.4998	$&$	5.7349	$&$	-2.1483	$&$	1.3	 \times 10^{-3}	$ \\
$	C14	$&$	3.4985	$&$	6.8369	$&$	1.5032	$&$	0.4999	$&$	5.7590	$&$	-2.1448	$&$	1.3	 \times 10^{-3}	$ \\
$	C15	$&$	3.5391	$&$	6.7190	$&$	1.5033	$&$	0.5000	$&$	5.7867	$&$	-2.1432	$&$	1.3	 \times 10^{-3}	$ \\
$	C16	$&$	3.5823	$&$	6.5967	$&$	1.5035	$&$	0.4998	$&$	5.8157	$&$	-2.1426	$&$	1.4	 \times 10^{-3}	$ \\
$	C17	$&$	3.6274	$&$	6.4733	$&$	1.5035	$&$	0.4998	$&$	5.8463	$&$	-2.1405	$&$	1.4	 \times 10^{-3}	$ \\
$	C18	$&$	3.6747	$&$	6.3467	$&$	1.5032	$&$	0.5000	$&$	5.8765	$&$	-2.1370	$&$	1.4	 \times 10^{-3}	$ \\
$	C19	$&$	3.7245	$&$	6.2193	$&$	1.5032	$&$	0.4999	$&$	5.9102	$&$	-2.1353	$&$	1.4	 \times 10^{-3}	$ \\
$	C20	$&$	3.7764	$&$	6.0914	$&$	1.5036	$&$	0.4998	$&$	5.9468	$&$	-2.1349	$&$	1.5	 \times 10^{-3}	$ \\
$	C21	$&$	3.8311	$&$	5.9605	$&$	1.5034	$&$	0.4998	$&$	5.9826	$&$	-2.1320	$&$	1.5	 \times 10^{-3}	$ \\
$	C22	$&$	3.8878	$&$	5.8291	$&$	1.5032	$&$	0.5001	$&$	6.0197	$&$	-2.1286	$&$	1.5	 \times 10^{-3}	$ \\
$	C23	$&$	3.9475	$&$	5.6967	$&$	1.5032	$&$	0.4999	$&$	6.0592	$&$	-2.1266	$&$	1.6	 \times 10^{-3}	$ \\
$	C24	$&$	4.0093	$&$	5.5658	$&$	1.5036	$&$	0.4998	$&$	6.1025	$&$	-2.1261	$&$	1.6	 \times 10^{-3}	$ \\
$	C25	$&$	4.0742	$&$	5.4324	$&$	1.5034	$&$	0.4999	$&$	6.1443	$&$	-2.1232	$&$	1.7	 \times 10^{-3}	$ \\
$	C26	$&$	4.1412	$&$	5.2995	$&$	1.5031	$&$	0.5001	$&$	6.1874	$&$	-2.1194	$&$	1.7	 \times 10^{-3}	$ \\
$	C27	$&$	4.2115	$&$	5.1674	$&$	1.5031	$&$	0.5000	$&$	6.2340	$&$	-2.1173	$&$	1.8	 \times 10^{-3}	$ \\
$	C28	$&$	4.2844	$&$	5.0359	$&$	1.5035	$&$	0.4999	$&$	6.2833	$&$	-2.1166	$&$	1.8	 \times 10^{-3}	$ \\
$	C29	$&$	4.3605	$&$	4.9039	$&$	1.5035	$&$	0.4999	$&$	6.3327	$&$	-2.1146	$&$	1.9	 \times 10^{-3}	$ \\
$	C30	$&$	4.4396	$&$	4.7723	$&$	1.5031	$&$	0.5001	$&$	6.3821	$&$	-2.1100	$&$	1.9	 \times 10^{-3}	$ \\
$	C31	$&$	4.5218	$&$	4.6425	$&$	1.5031	$&$	0.5001	$&$	6.4355	$&$	-2.1076	$&$	2.0	 \times 10^{-3}	$ \\
$	C32	$&$	4.6077	$&$	4.5130	$&$	1.5033	$&$	0.5000	$&$	6.4912	$&$	-2.1066	$&$	2.0	 \times 10^{-3}	$ \\
$	C33	$&$	4.6142	$&$	4.5018	$&$	1.5040	$&$	0.4996	$&$	6.4939	$&$	-2.1098	$&$	2.0	 \times 10^{-3}	$ \\
$	C34	$&$	4.7030	$&$	4.3735	$&$	1.5033	$&$	0.5001	$&$	6.5480	$&$	-2.1038	$&$	2.1	 \times 10^{-3}	$ \\
$	C35	$&$	4.7960	$&$	4.2462	$&$	1.5030	$&$	0.5001	$&$	6.6056	$&$	-2.1003	$&$	2.2	 \times 10^{-3}	$ \\
$	C36	$&$	4.8938	$&$	4.1186	$&$	1.5027	$&$	0.5000	$&$	6.6642	$&$	-2.0972	$&$	2.2	 \times 10^{-3}	$ \\
           \enddata
\end{deluxetable}
\clearpage

\begin{landscape}
\begin{deluxetable}{rrrrrrrrrrr}
   \tablenum{9}
   \tablewidth{0pt}
   \tablecolumns{11}
   \tablecaption{Individual Stellar Components along DWD Inspiral
                 Sequence `C'\label{tab:scf_models_ind_q_0.50}}
   \tablehead{
     \colhead{Model} &
           \colhead{$M_1$} &
           \colhead{$R_1$}  &
           \colhead{$\rho^{i=1}_\mathrm{max}$}  &
           \colhead{$f^{i=1}_\mathrm{RL}$} &
           \colhead{$J^{i=1}_\mathrm{spin}$} &
           \colhead{$M_2$} &
           \colhead{$R_2$}  &
           \colhead{$\rho^{i=2}_\mathrm{max}$}  &
           \colhead{$f^{i=2}_\mathrm{RL}$} &
           \colhead{$J^{i=2}_\mathrm{spin}$} \\
            &
           \colhead{($M_{\sun}$)} &
           \colhead{($10^9~\mathrm{cm}$)} &
           \colhead{($10^9~\mathrm{cgs}$)} &
           &
           \colhead{($10^{50}~\mathrm{cgs}$)} &
           \colhead{($M_{\sun}$)} &
           \colhead{($10^9~\mathrm{cm}$)} &
           \colhead{($10^9~\mathrm{cgs}$)} &
           &
           \colhead{($10^{50}~\mathrm{cgs}$)}
           }

           \startdata
$	C1	$&$	1.0028	$&$	0.5542	$&$	31.824	$&$	0.0627	$&$	0.0800	$&$	0.5015	$&$	1.0110	$&$	1.767	$&$	1.0000	$&$	0.1613	$ \\
$	C2	$&$	1.0029	$&$	0.5541	$&$	31.824	$&$	0.0626	$&$	0.0799	$&$	0.5015	$&$	1.0108	$&$	1.767	$&$	0.9987	$&$	0.1613	$ \\
$	C3	$&$	1.0025	$&$	0.5544	$&$	31.824	$&$	0.0624	$&$	0.0798	$&$	0.5014	$&$	1.0104	$&$	1.767	$&$	0.9927	$&$	0.1607	$ \\
$	C4	$&$	1.0025	$&$	0.5546	$&$	31.824	$&$	0.0612	$&$	0.0790	$&$	0.5009	$&$	1.0093	$&$	1.767	$&$	0.9703	$&$	0.1586	$ \\
$	C5	$&$	1.0027	$&$	0.5541	$&$	31.806	$&$	0.0602	$&$	0.0784	$&$	0.5010	$&$	1.0081	$&$	1.772	$&$	0.9536	$&$	0.1571	$ \\
$	C6	$&$	1.0025	$&$	0.5543	$&$	31.812	$&$	0.0593	$&$	0.0778	$&$	0.5011	$&$	1.0068	$&$	1.776	$&$	0.9334	$&$	0.1553	$ \\
$	C7	$&$	1.0023	$&$	0.5547	$&$	31.834	$&$	0.0583	$&$	0.0771	$&$	0.5011	$&$	1.0051	$&$	1.780	$&$	0.9109	$&$	0.1534	$ \\
$	C8	$&$	1.0024	$&$	0.5541	$&$	31.845	$&$	0.0568	$&$	0.0762	$&$	0.5010	$&$	1.0035	$&$	1.786	$&$	0.8873	$&$	0.1512	$ \\
$	C9	$&$	1.0028	$&$	0.5540	$&$	31.823	$&$	0.0554	$&$	0.0753	$&$	0.5010	$&$	1.0021	$&$	1.791	$&$	0.8634	$&$	0.1489	$ \\
$	C10	$&$	1.0023	$&$	0.5544	$&$	31.847	$&$	0.0542	$&$	0.0744	$&$	0.5010	$&$	1.0007	$&$	1.796	$&$	0.8375	$&$	0.1464	$ \\
$	C11	$&$	1.0022	$&$	0.5541	$&$	31.878	$&$	0.0527	$&$	0.0734	$&$	0.5010	$&$	0.9993	$&$	1.801	$&$	0.8112	$&$	0.1438	$ \\
$	C12	$&$	1.0024	$&$	0.5537	$&$	31.901	$&$	0.0497	$&$	0.0713	$&$	0.5011	$&$	0.9961	$&$	1.814	$&$	0.7593	$&$	0.1388	$ \\
$	C13	$&$	1.0025	$&$	0.5537	$&$	31.901	$&$	0.0480	$&$	0.0701	$&$	0.5010	$&$	0.9943	$&$	1.819	$&$	0.7309	$&$	0.1360	$ \\
$	C14	$&$	1.0022	$&$	0.5539	$&$	31.934	$&$	0.0465	$&$	0.0689	$&$	0.5010	$&$	0.9933	$&$	1.824	$&$	0.7048	$&$	0.1332	$ \\
$	C15	$&$	1.0022	$&$	0.5537	$&$	31.965	$&$	0.0448	$&$	0.0677	$&$	0.5011	$&$	0.9919	$&$	1.829	$&$	0.6777	$&$	0.1303	$ \\
$	C16	$&$	1.0025	$&$	0.5533	$&$	31.971	$&$	0.0431	$&$	0.0664	$&$	0.5010	$&$	0.9905	$&$	1.836	$&$	0.6507	$&$	0.1274	$ \\
$	C17	$&$	1.0024	$&$	0.5533	$&$	31.982	$&$	0.0415	$&$	0.0651	$&$	0.5010	$&$	0.9887	$&$	1.841	$&$	0.6239	$&$	0.1245	$ \\
$	C18	$&$	1.0021	$&$	0.5534	$&$	32.020	$&$	0.0399	$&$	0.0639	$&$	0.5010	$&$	0.9873	$&$	1.847	$&$	0.5983	$&$	0.1216	$ \\
$	C19	$&$	1.0022	$&$	0.5532	$&$	32.053	$&$	0.0383	$&$	0.0625	$&$	0.5010	$&$	0.9861	$&$	1.852	$&$	0.5728	$&$	0.1187	$ \\
$	C20	$&$	1.0025	$&$	0.5530	$&$	32.057	$&$	0.0366	$&$	0.0612	$&$	0.5011	$&$	0.9850	$&$	1.857	$&$	0.5488	$&$	0.1158	$ \\
$	C21	$&$	1.0024	$&$	0.5530	$&$	32.072	$&$	0.0351	$&$	0.0598	$&$	0.5010	$&$	0.9834	$&$	1.863	$&$	0.5242	$&$	0.1129	$ \\
$	C22	$&$	1.0021	$&$	0.5525	$&$	32.115	$&$	0.0335	$&$	0.0585	$&$	0.5011	$&$	0.9827	$&$	1.867	$&$	0.5012	$&$	0.1100	$ \\
$	C23	$&$	1.0021	$&$	0.5530	$&$	32.152	$&$	0.0320	$&$	0.0572	$&$	0.5010	$&$	0.9815	$&$	1.873	$&$	0.4788	$&$	0.1072	$ \\
$	C24	$&$	1.0025	$&$	0.5528	$&$	32.158	$&$	0.0305	$&$	0.0558	$&$	0.5011	$&$	0.9804	$&$	1.877	$&$	0.4571	$&$	0.1044	$ \\
$	C25	$&$	1.0024	$&$	0.5522	$&$	32.175	$&$	0.0290	$&$	0.0544	$&$	0.5010	$&$	0.9793	$&$	1.882	$&$	0.4357	$&$	0.1015	$ \\
$	C26	$&$	1.0020	$&$	0.5524	$&$	32.223	$&$	0.0276	$&$	0.0531	$&$	0.5011	$&$	0.9786	$&$	1.886	$&$	0.4154	$&$	0.0988	$ \\
$	C27	$&$	1.0021	$&$	0.5523	$&$	32.267	$&$	0.0262	$&$	0.0517	$&$	0.5011	$&$	0.9775	$&$	1.891	$&$	0.3951	$&$	0.0960	$ \\
$	C28	$&$	1.0024	$&$	0.5515	$&$	32.281	$&$	0.0248	$&$	0.0503	$&$	0.5011	$&$	0.9761	$&$	1.895	$&$	0.3755	$&$	0.0933	$ \\
$	C29	$&$	1.0024	$&$	0.5515	$&$	32.291	$&$	0.0235	$&$	0.0489	$&$	0.5011	$&$	0.9753	$&$	1.899	$&$	0.3572	$&$	0.0906	$ \\
$	C30	$&$	1.0020	$&$	0.5518	$&$	32.342	$&$	0.0223	$&$	0.0476	$&$	0.5011	$&$	0.9746	$&$	1.903	$&$	0.3393	$&$	0.0879	$ \\
$	C31	$&$	1.0020	$&$	0.5516	$&$	32.397	$&$	0.0211	$&$	0.0463	$&$	0.5011	$&$	0.9739	$&$	1.907	$&$	0.3219	$&$	0.0853	$ \\
$	C32	$&$	1.0022	$&$	0.5510	$&$	32.425	$&$	0.0199	$&$	0.0450	$&$	0.5011	$&$	0.9727	$&$	1.911	$&$	0.3049	$&$	0.0827	$ \\
$	C33	$&$	1.0029	$&$	0.5505	$&$	32.425	$&$	0.0197	$&$	0.0448	$&$	0.5011	$&$	0.9727	$&$	1.911	$&$	0.3041	$&$	0.0825	$ \\
$	C34	$&$	1.0021	$&$	0.5511	$&$	32.462	$&$	0.0187	$&$	0.0435	$&$	0.5011	$&$	0.9722	$&$	1.915	$&$	0.2883	$&$	0.0800	$ \\
$	C35	$&$	1.0019	$&$	0.5508	$&$	32.522	$&$	0.0176	$&$	0.0422	$&$	0.5011	$&$	0.9720	$&$	1.919	$&$	0.2735	$&$	0.0775	$ \\
$	C36	$&$	1.0018	$&$	0.5508	$&$	32.522	$&$	0.0166	$&$	0.0409	$&$	0.5009	$&$	0.9711	$&$	1.919	$&$	0.2582	$&$	0.0750	$ \\
           \enddata
\end{deluxetable}
\end{landscape}

\begin{deluxetable}{rrrrrrrr}
   \tablenum{10}
   \tablewidth{0pt}
   \tablecolumns{7}
   \tablecaption{Semi-detached DWD Sequence `D'; $M_\mathrm{tot}=1.5 M_\odot$ \label{tab:contactD}}
   \tablehead{
           \colhead{$M_\mathrm{tot}$}  &
           \colhead{$q$} &
           \colhead{$a$} &
           \colhead{$\Omega$}  &
           \colhead{$J_\mathrm{tot}$} &
           \colhead{$E_\mathrm{tot}$}  &
           \colhead{VE} \\
           \colhead{($M_\odot$)}&  &
           \colhead{($10^9$ cm)} &
           \colhead{($10^{-2}~\mathrm{s}^{-1}$)} &
           \colhead{($10^{50}~\mathrm{cgs}$)} &
           \colhead{($10^{50}~\mathrm{erg}$)} &
           }

           \startdata
$	1.5048	$&$	1.0000	$&$	2.1010	$&$	14.8472	$&$	5.3866	$&$	-1.8636	$&$	1.5	 \times 10^{-4}	$ \\
$	1.5034	$&$	0.9508	$&$	2.1638	$&$	14.1787	$&$	5.4116	$&$	-1.8566	$&$	1.8	 \times 10^{-4}	$ \\
$	1.5045	$&$	0.8994	$&$	2.2336	$&$	13.4444	$&$	5.4491	$&$	-1.8611	$&$	9.5	 \times 10^{-5}	$ \\
$	1.5034	$&$	0.8507	$&$	2.3102	$&$	12.8287	$&$	5.4797	$&$	-1.8628	$&$	1.4	 \times 10^{-4}	$ \\
$	1.5033	$&$	0.8007	$&$	2.3961	$&$	12.1414	$&$	5.5141	$&$	-1.8736	$&$	2.2	 \times 10^{-4}	$ \\
$	1.5034	$&$	0.7504	$&$	2.4919	$&$	11.4431	$&$	5.5460	$&$	-1.8925	$&$	3.1	 \times 10^{-4}	$ \\
$	1.5034	$&$	0.7004	$&$	2.5987	$&$	10.7442	$&$	5.5702	$&$	-1.9194	$&$	4.0	 \times 10^{-4}	$ \\
$	1.5034	$&$	0.6504	$&$	2.7188	$&$	10.0382	$&$	5.5871	$&$	-1.9572	$&$	5.5	 \times 10^{-4}	$ \\
$	1.5032	$&$	0.6004	$&$	2.8550	$&$	9.3217	$&$	5.5914	$&$	-2.0072	$&$	7.2	 \times 10^{-4}	$ \\
$	1.5033	$&$	0.5503	$&$	3.0109	$&$	9.0670	$&$	5.5804	$&$	-2.0741	$&$	9.1	 \times 10^{-4}	$ \\
$	1.5031	$&$	0.5004	$&$	3.1914	$&$	7.8907	$&$	5.5484	$&$	
-2.1587	$&$	1.2	 \times 10^{-3}	$
           \enddata
\end{deluxetable}

\begin{deluxetable}{rrrrrrrr}
   \tablenum{11}
   \tablewidth{0pt}
   \tablecolumns{7}
   \tablecaption{Semi-detached DWD Sequence `E'; $M_\mathrm{tot}=1.0 M_\odot$ \label{tab:contactE}}
   \tablehead{
           \colhead{$M_\mathrm{tot}$}  &
           \colhead{$q$} &
           \colhead{$a$} &
           \colhead{$\Omega$}  &
           \colhead{$J_\mathrm{tot}$} &
           \colhead{$E_\mathrm{tot}$}  &
           \colhead{VE} \\
           \colhead{($M_\odot$)}&  &
           \colhead{($10^9$ cm)} &
           \colhead{($10^{-2}~\mathrm{s}^{-1}$)} &
           \colhead{($10^{50}~\mathrm{cgs}$)} &
           \colhead{($10^{50}~\mathrm{erg}$)} &
           }

           \startdata
$	1.0030	$&$	1.0000	$&$	2.7109	$&$	8.2477	$&$	3.3579	$&$	-0.6657	$&$	1.7	 \times 10^{-4}	$ \\
$	1.0024	$&$	0.9504	$&$	2.7778	$&$	7.9369	$&$	3.3687	$&$	-0.6642	$&$	1.6	 \times 10^{-4}	$ \\
$	1.0019	$&$	0.9001	$&$	2.8641	$&$	7.5710	$&$	3.3850	$&$	-0.6636	$&$	2.1	 \times 10^{-4}	$ \\
$	1.0022	$&$	0.8508	$&$	2.9380	$&$	7.2829	$&$	3.3983	$&$	-0.6659	$&$	2.2	 \times 10^{-4}	$ \\
$	1.0024	$&$	0.8003	$&$	3.0161	$&$	6.9993	$&$	3.4073	$&$	-0.6698	$&$	2.2	 \times 10^{-4}	$ \\
$	1.0024	$&$	0.7503	$&$	3.1165	$&$	6.6609	$&$	3.4163	$&$	-0.6751	$&$	2.6	 \times 10^{-4}	$ \\
$	1.0024	$&$	0.7003	$&$	3.2294	$&$	6.3129	$&$	3.4212	$&$	-0.6828	$&$	3.3	 \times 10^{-4}	$ \\
$	1.0023	$&$	0.6502	$&$	3.3569	$&$	5.9551	$&$	3.4204	$&$	-0.6934	$&$	3.9	 \times 10^{-4}	$ \\
$	1.0023	$&$	0.6003	$&$	3.5026	$&$	5.5862	$&$	3.4123	$&$	-0.7074	$&$	4.7	 \times 10^{-4}	$ \\
$	1.0022	$&$	0.5503	$&$	3.6696	$&$	5.2078	$&$	3.3943	$&$	-0.7254	$&$	5.6	 \times 10^{-4}	$ \\
$	1.0021	$&$	0.5003	$&$	3.8649	$&$	4.8169	$&$	3.3644	$&$	-0.7484	$&$	6.8	 \times 10^{-4}	$ \\
$	1.0020	$&$	0.4504	$&$	4.0955	$&$	4.4134	$&$	3.3184	$&$	-0.7773	$&$	8.0	 \times 10^{-4}	$ \\
           \enddata
\end{deluxetable}
\clearpage

\begin{deluxetable}{lcc}
    \tablewidth{0pt}
    \tablenum{A1}
    \tablecolumns{5}
    \tablecaption{Physical Constants \label{tab:physicalConstants}}
    \tablehead{
      \colhead{Constants\tablenotemark{a}} & \colhead{This Paper\tablenotemark{b}} & \colhead{\cite{Chandrasekhar67}\tablenotemark{c}} \\
                 \colhead{(1)} & \colhead{(2)} & \colhead{(3)}}

    \startdata
      $c$ ($\mathrm{cm}~\mathrm{s}^{-1}$)        &  $2.99792\times 10^{+10}$ & $2.9978\times 10^{10}$   \\
      $h$ ($\mathrm{erg}\cdot\mathrm{s}$)        &  $6.62608\times 10^{-27}$ & $6.62\times 10^{-27}$    \\
      $m_e$ (g)                                  &  $9.10939\times 10^{-28}$ & $9.105\times 10^{-28}$   \\
      $m_p$ (g)                                  &  $1.67262\times 10^{-24}$   & $1.672\times 10^{-24}$   \\
      $m_\mu$   (g)                              &  $1.66054\times 10^{-24}$   & $\cdots$   \\
      $G$ ($\mathrm{cm}^3 ~\mathrm{g}^{-1}~\mathrm{s}^{-2}$)
                                                 &  $6.6726\times 10^{-8}$  & $6.62\times 10^{-8}$    \\
      $M_\odot$ (g)                              &  $1.9891 \times 10^{33}$    & $1.985\times 10^{33}$    \\
      $R_\odot$ (cm)                             &  $6.955\times 10^{10}$     & $6.951\times
      10^{10}$ \\
      $A$ (dynes $\mathrm{cm}^{-2}$)      & $6.00228\times 10^{22}$    &$6.01\times 10^{22}$ \\
      $B\mu_e^{-1}$ (g $\mathrm{cm}^{-3}$)      & $9.81011\times 10^5$     &$9.82\times 10^5$ \\
      $\ell_1\mu_e$ (cm)      & $7.71395\times 10^8$     &$7.705\times 10^8$
    \enddata
\tablenotetext{a}{Speed of light, $c$; Planck's constant, $h$;
mass of the electron, $m_e$; mass of the proton $m_p$; atomic
mass unit, $m_\mu$; universal gravitational constant, $G$;
solar mass, $M_\odot$; solar radius, $R_\odot$; as used in the
ZTWD equation of state (\ref{eq:EOSchandra}), $A = \pi m_e^4
c^5/3h^3$ and $B\mu_e^{-1} =8\pi m_e^3 c^3 m_p/3h^3$; the
characteristic WD length scale, $\ell_1 \mu_e = (2A/\pi
G)^{1/2} (\mu_e/B)$.} \tablenotetext{b}{Drawn from
\cite{Cox99}.} \tablenotetext{c}{Drawn from Appendix I, Table
32 of \cite{Chandrasekhar67}.}
\end{deluxetable}

\newpage


\begin{landscape}

\newpage
\begin{figure}[ht]
\epsfig{file=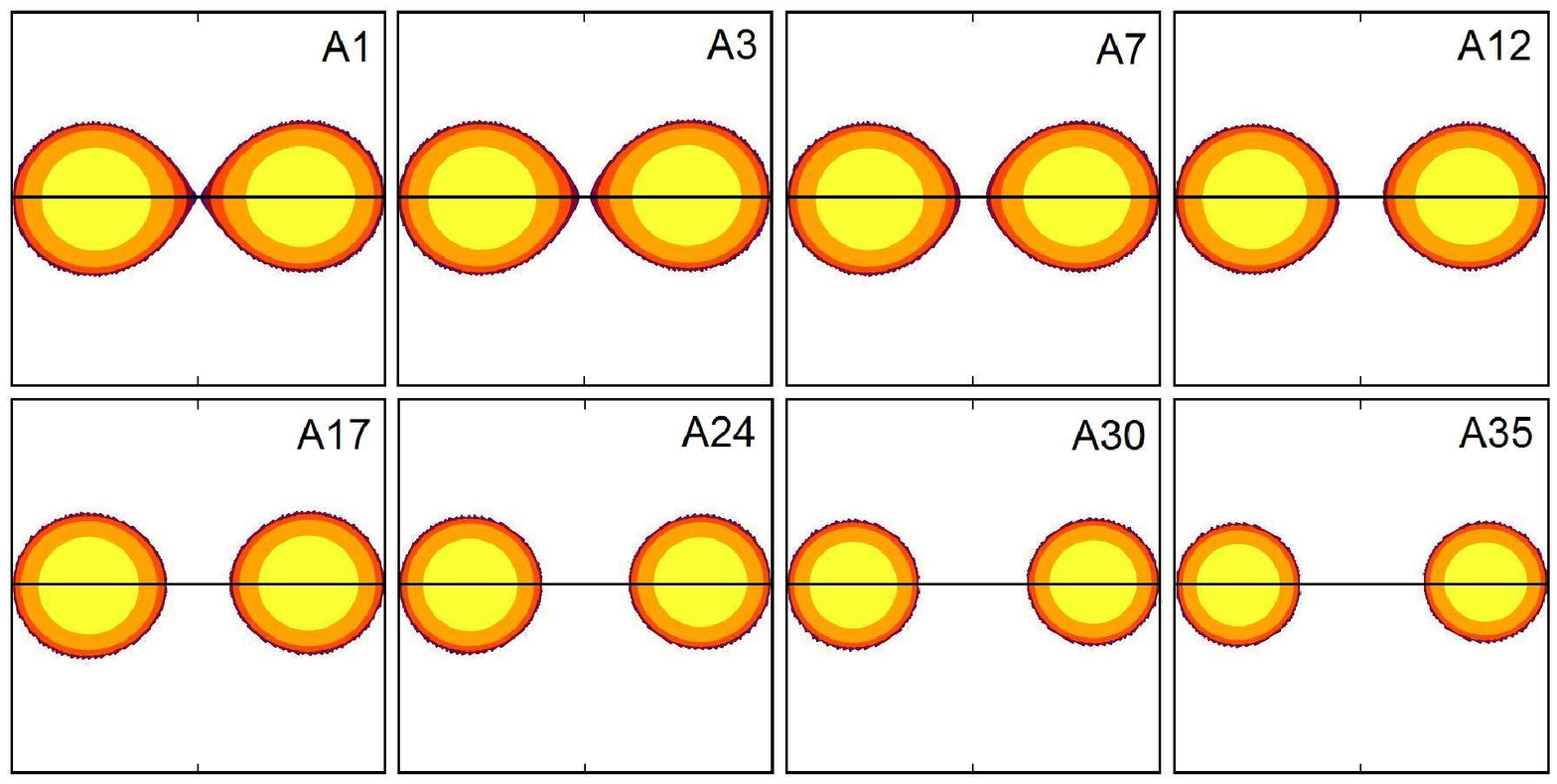,width=\textwidth}
\figcaption[f3.ps]{Density contours in the equatorial plane
are shown for eight separate ZTWD binary models with increasing separation
along inspiral sequence `A' ($M_\mathrm{tot} = 1.5 M_\odot$; $q=1$).
Labels in the upper-right-hand corner of each panel identify each model
by number according to its corresponding position along the sequence as
itemized in Tables \ref{tab:scf_models_q_1.00} and
\ref{tab:scf_models_ind_q_1.00}.\label{fig:DenSeries100}}
\end{figure}


\newpage
\begin{figure}[ht]
\epsfig{file=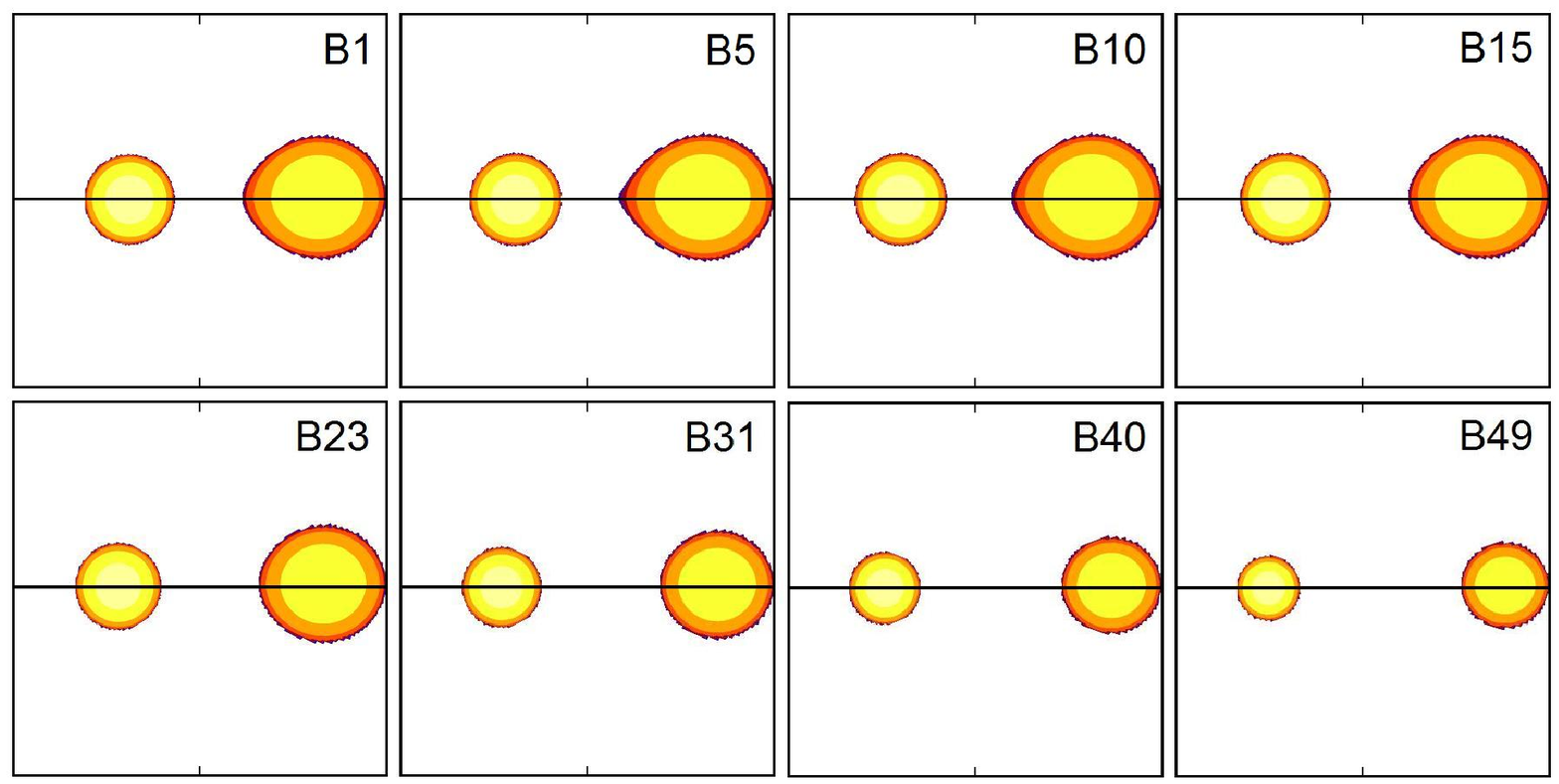,width=\textwidth}
\figcaption[f4.ps]{Density contours in the equatorial plane
are shown for eight separate ZTWD binary models with increasing separation
along inspiral sequence `B' ($M_\mathrm{tot} = 1.5 M_\odot$; $q=2/3$).
Labels in the upper-right-hand corner of each panel identify each model
by number according to its corresponding position along the sequence as
itemized in Tables \ref{tab:scf_models_q_0.66} and
\ref{tab:scf_models_ind_q_0.66}.\label{fig:DenSeries066}}
\end{figure}


\newpage
\begin{figure}[ht]
\epsfig{file=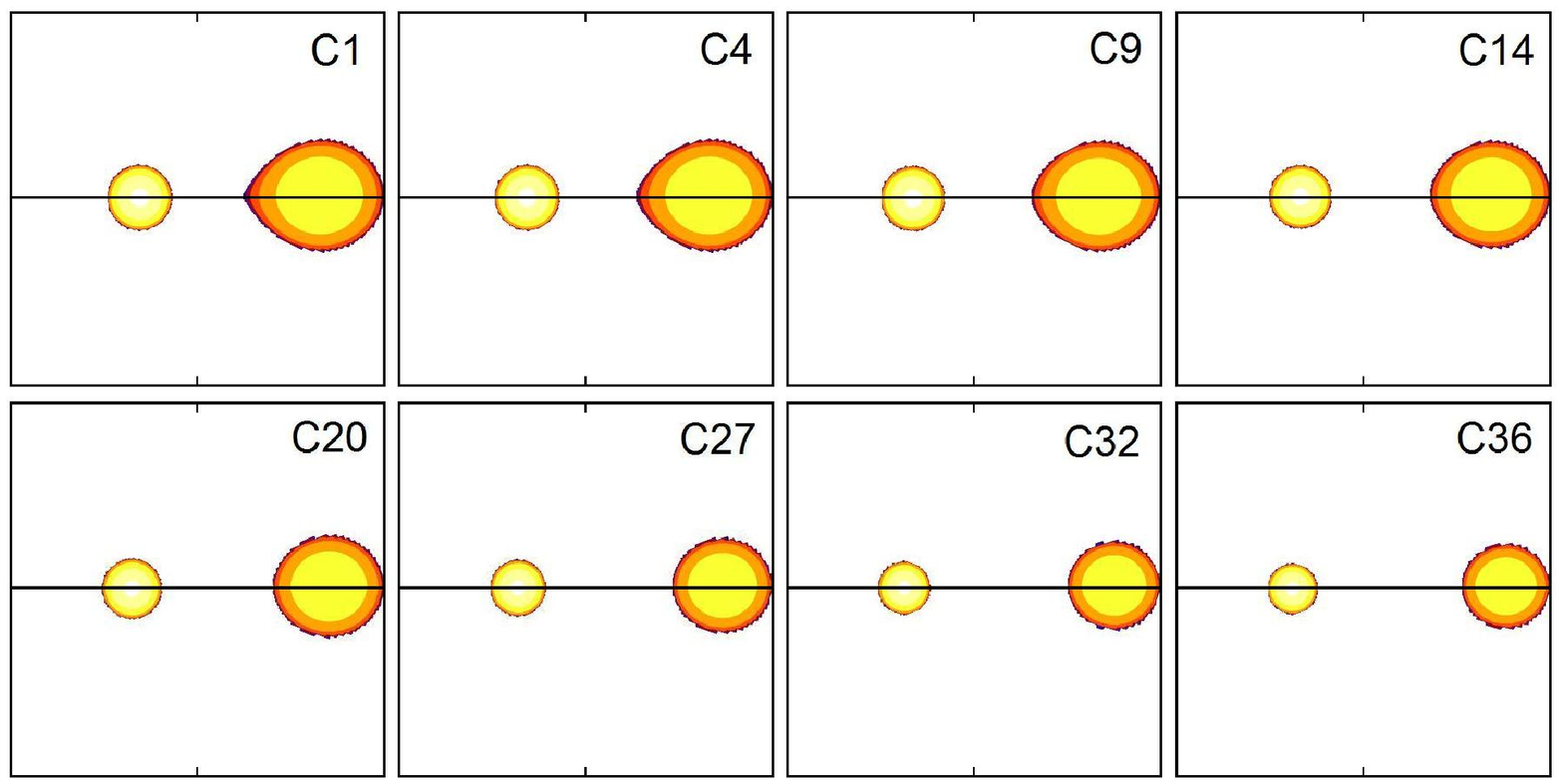,width=\textwidth}
\figcaption[f5.ps]{Density contours in the equatorial plane
are shown for eight separate ZTWD binary models with increasing separation
along inspiral sequence `C' ($M_\mathrm{tot} = 1.5 M_\odot$; $q=1/2$).
Labels in the upper-right-hand corner of each panel identify each model
by number according to its corresponding position along the sequence as
itemized in Tables \ref{tab:scf_models_q_0.50} and
\ref{tab:scf_models_ind_q_0.50}.\label{fig:DenSeries050}}
\end{figure}


\end{landscape}

\newpage
\begin{figure}[ht]
\epsfig{file=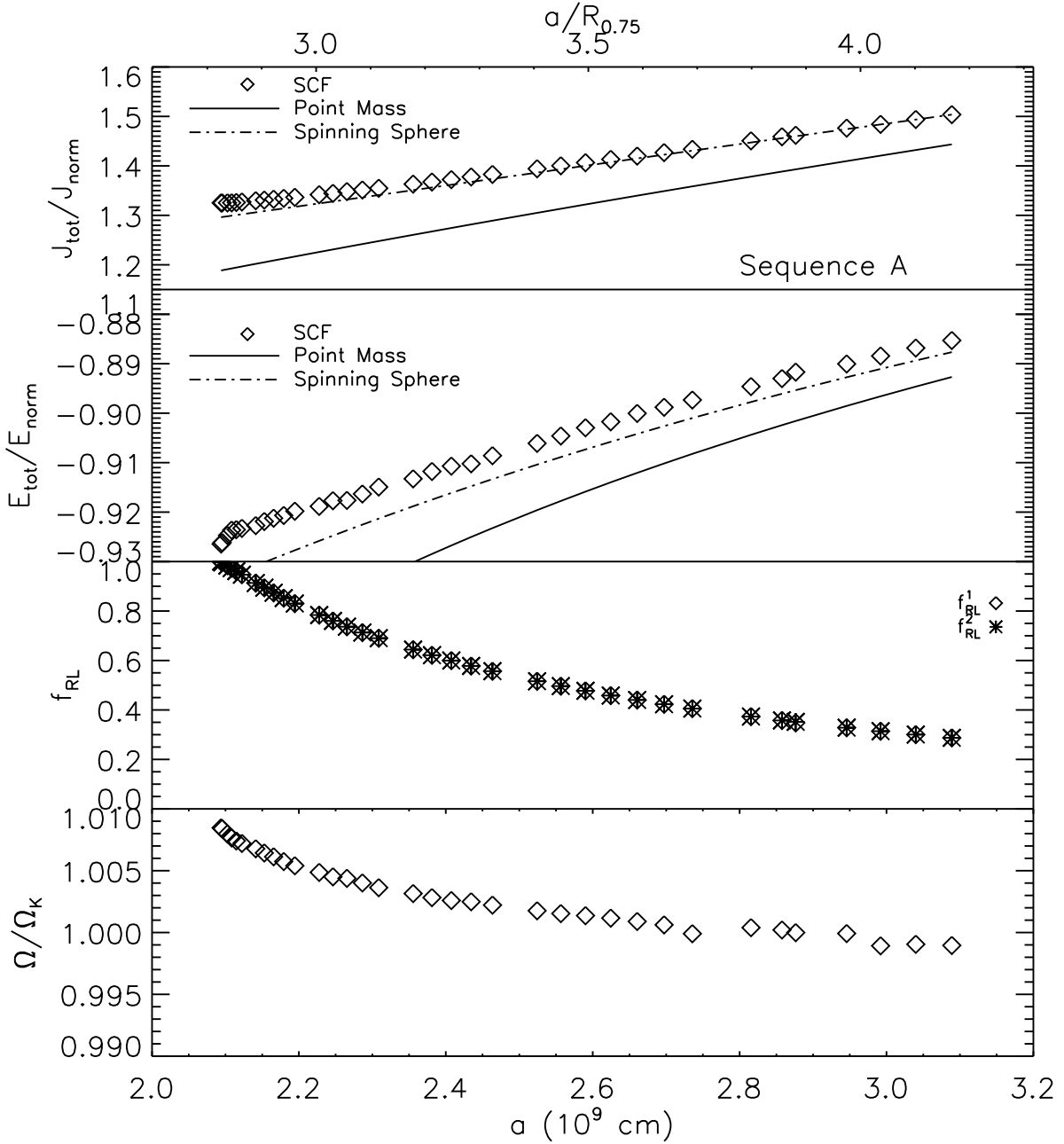,width=\textwidth}
\figcaption[f6.ps]{({\it Top panel}) Normalized
total angular momentum, $J_\mathrm{tot}/J_\mathrm{norm}$, ({\it
second panel}) normalized total energy,
$E_\mathrm{tot}/E_\mathrm{norm}$, ({\it third panel}) the
Roche-lobe filling factor, $f_\mathrm{RL}$, for the secondary
(asterisks) and primary (diamonds) stars, and ({\it bottom
panel}) the normalized orbital angular velocity,
$\Omega/\Omega_\mathrm{K}$, are plotted as a function of binary
separation for models $A1$ through $A35$ along inspiral
sequence `A' ($M_\mathrm{tot}=1.5 M_\odot$; $q=1$). Data for
the individual models is drawn from Tables
\ref{tab:scf_models_q_1.00}  and
\ref{tab:scf_models_ind_q_1.00}; the separation $a$ is labeled
in units of $10^9~\mathrm{cm}$ along the bottom axis and as a
ratio to $R_{0.75}$ along the top axis.  The solid curves in
the top two panels display the analytic functions
$J_\mathrm{pm}(a)/J_\mathrm{norm}$ and
$E_\mathrm{pm}(a)/E_\mathrm{norm}$ given, respectively, by
Eqs.~(\ref{eq:Jpm}) and (\ref{eq:Epm}) for a ``point-mass''
sequence of the specified total mass and mass ratio; and the
dot-dashed curves display the analytic functions
$J_\mathrm{ss}(a)/J_\mathrm{norm}$ and
$E_\mathrm{ss}(a)/E_\mathrm{norm}$ appropriate for a ``spinning
spheres'' sequence given, respectively, by Eqs.~(\ref{eq:Jss})
and (\ref{eq:Ess}).\label{fig:075_075series}}
\end{figure}


\begin{figure}[ht]
\epsfig{file=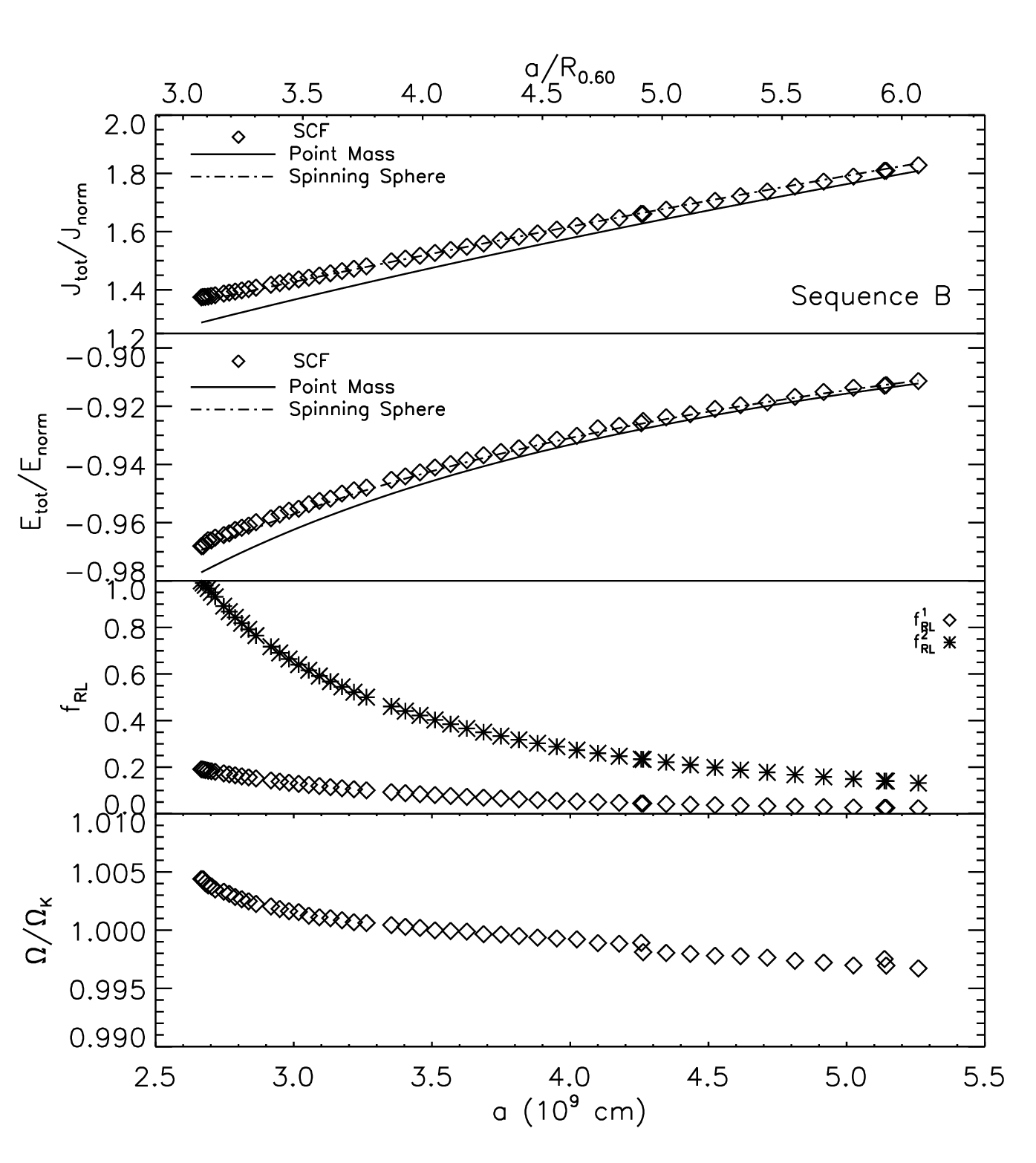,width=\textwidth}
\figcaption[f7.ps]{Same as
Fig.~\ref{fig:075_075series} but for models $B1$ through $B49$
along the inspiral sequence `B' ($M_\mathrm{tot}=1.5 M_\odot$;
$q=2/3$), as tabulated in Tables \ref{tab:scf_models_q_0.66}
and \ref{tab:scf_models_ind_q_0.66}; along the top axis, the
separation $a$ is labeled as a ratio to $R_{0.60}$.
\label{fig:090_060series}}
\end{figure}


\begin{figure}[ht]
\epsfig{file=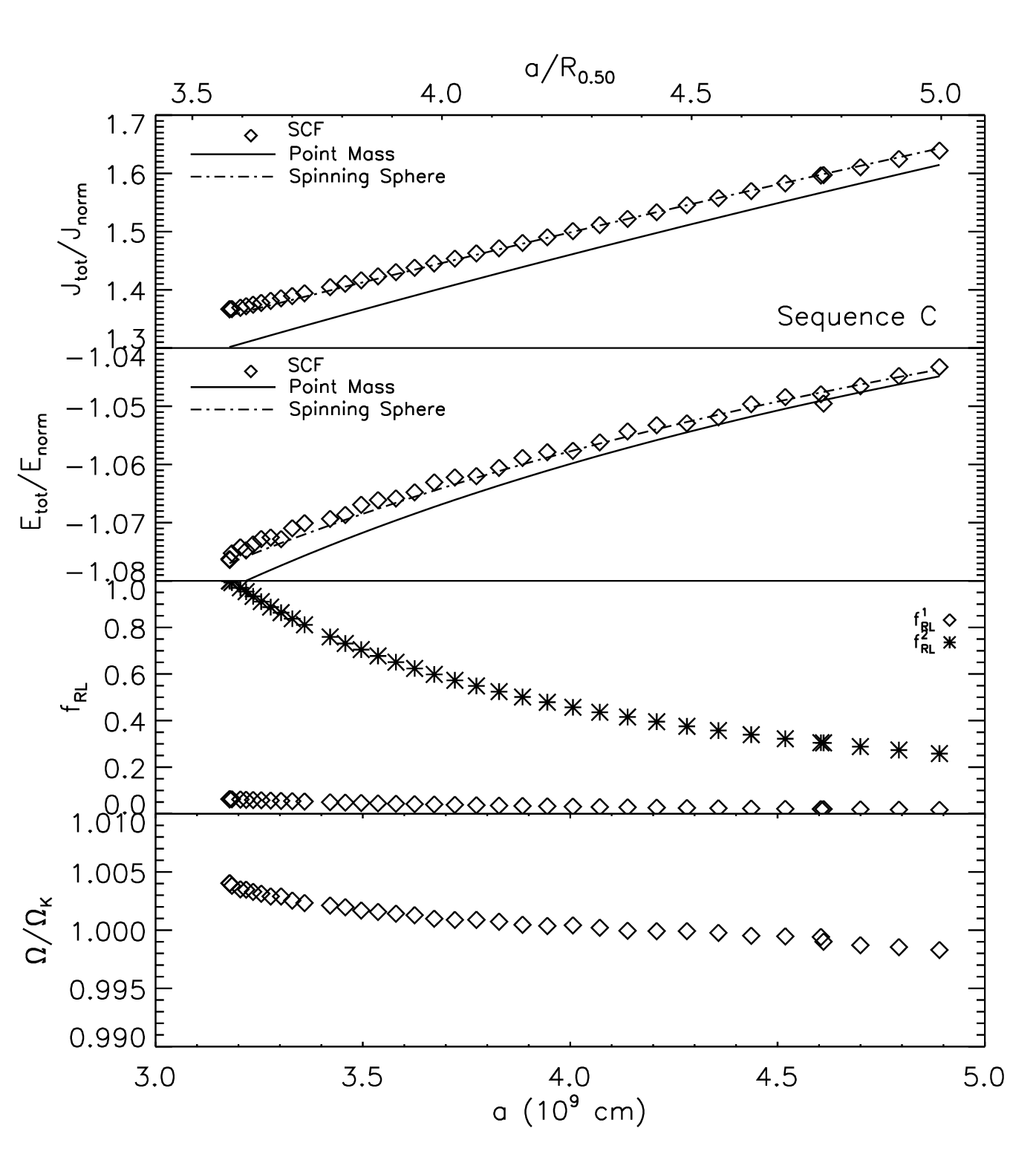,width=\textwidth}
\figcaption[f8.ps] {Same as
Fig.~\ref{fig:075_075series} but for models $C1$ through $C36$
along the inspiral sequence `C' ($M_\mathrm{tot}=1.5 M_\odot$;
$q=1/2$), as tabulated in Tables \ref{tab:scf_models_q_0.50}
and \ref{tab:scf_models_ind_q_0.50}; along the top axis, the
separation $a$ is labeled as a ratio to $R_{0.50}$.
\label{fig:100_050series}}
\end{figure}


\begin{figure}[ht]
\epsfig{file=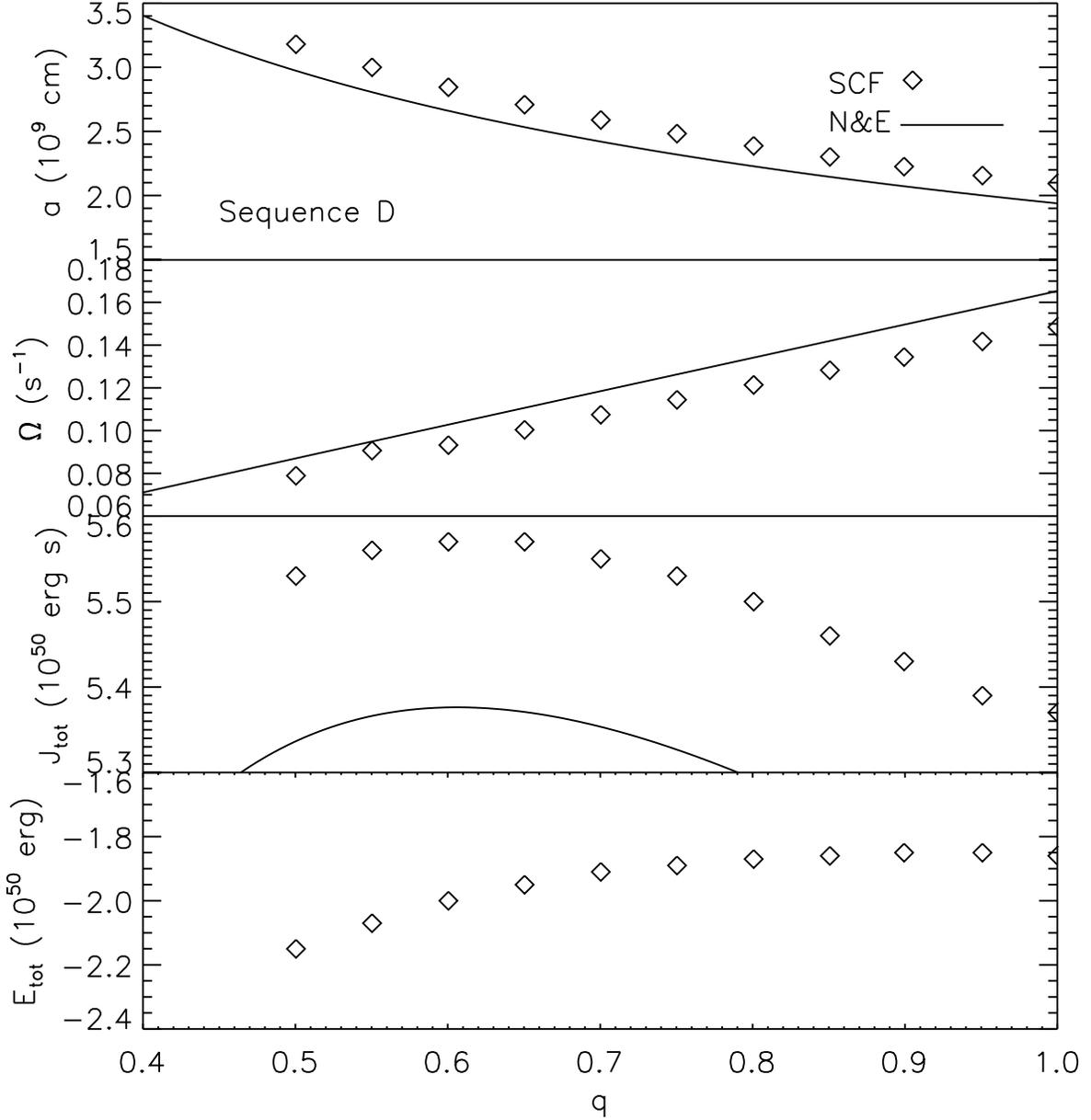,width=\textwidth}
\figcaption[f9.ps] {System parameters $a$, $\Omega$,
$J_\mathrm{tot}$, and $E_\mathrm{tot}$ at contact as a function
of mass ratio, $q$, for DWD systems having a total mass of $1.5
M_\odot$. Solid curves in the top three panels show predicted
behavior based on Nauenberg's (1972) and Eggleton's (1983)
approximate, analytic expressions as discussed in the text.
\label{fig:contactM1.5}}
\end{figure}


\begin{figure}[ht]
\epsfig{file=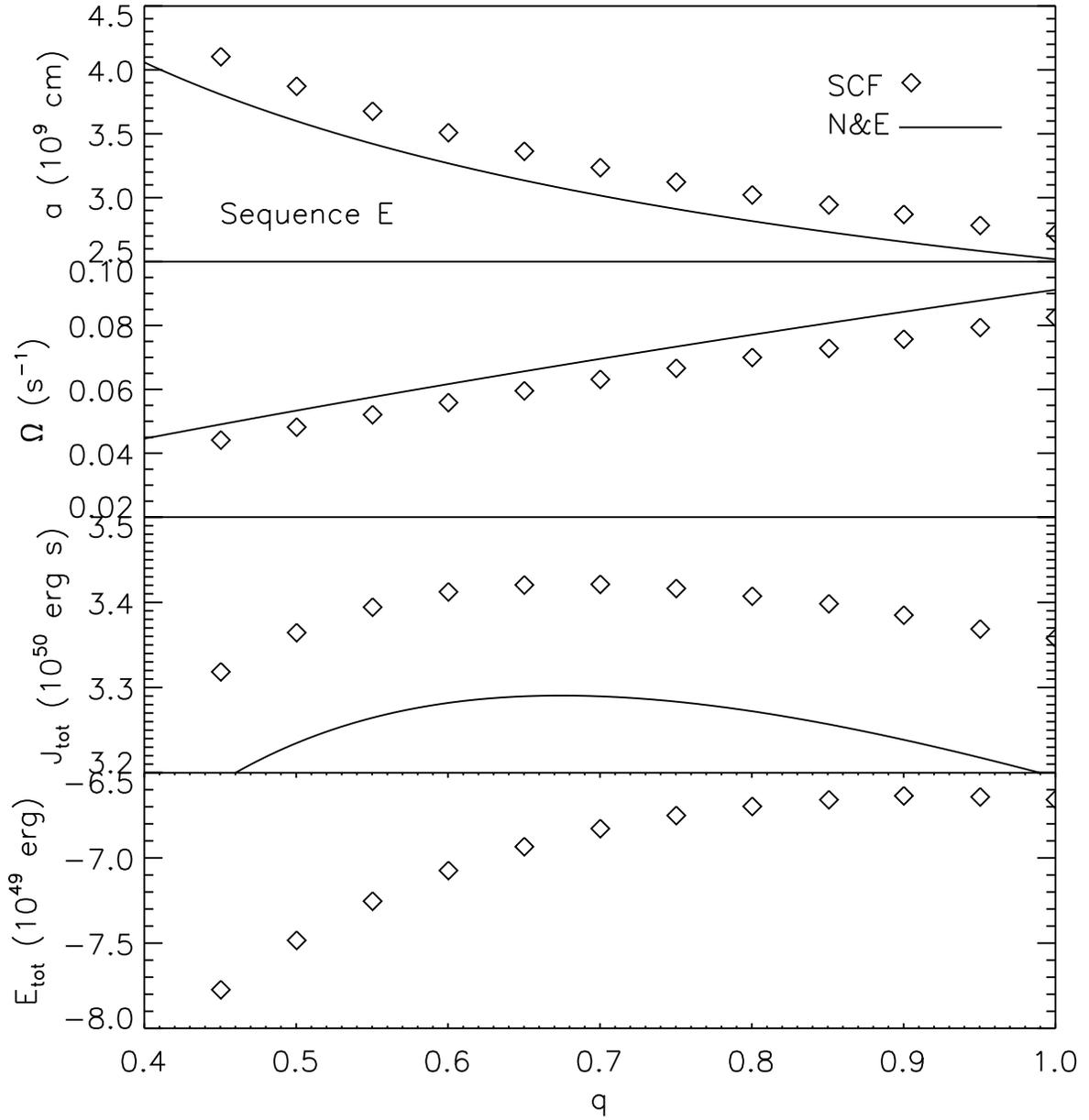,width=\textwidth}
\figcaption[f10.ps] {Same as Figure \ref{fig:contactM1.5},
but for DWD systems having a total mass of $1.0 M_\odot$.
\label{fig:contactM1.0}}
\end{figure}


\end{document}